\documentclass[reprint,superscriptaddress,floatfix,amsmath,amssymb,pra,aps,showpacs,twocolumn]{revtex4-1}
\usepackage{tabularx} 
\usepackage{amsmath,amsfonts,amssymb,amsthm,graphics,graphicx,epsfig,bbm}
\usepackage[colorlinks=true,citecolor=blue,linkcolor=blue,urlcolor=blue]{hyperref}
\usepackage[usenames]{color}
\usepackage{graphicx}
\usepackage{subfigure}
\usepackage{amsmath}
\usepackage{epsfig}
\usepackage{dcolumn}
\usepackage{bm}
\usepackage{color}
\usepackage{times}
\usepackage{appendix}
\usepackage{epstopdf}
\usepackage{amssymb}
\usepackage{amstext}
\usepackage{latexsym}
\usepackage{hyperref}
\usepackage{amsfonts}
\usepackage{psfrag}
\usepackage{float}
\usepackage{enumerate}
\usepackage[table,xcdraw]{xcolor}
\usepackage{longtable}
\usepackage{bm}

\newcommand{\ket}[1]{\vert #1 \rangle}
\newcommand{\bra}[1]{\langle #1 \vert}
\newcommand{\ketbra}[2]{\vert #1 \rangle \langle #2 \vert}
\newcommand{\braket}[2]{\langle #1 \vert #2 \rangle}

\newcommand{\abs}[1]{| #1 |}

\begin{document}
\title{Adaptive Random Quantum Eigensolver} 

\date{\today}
	
\author{N. Barraza}
\affiliation{International Center of Quantum Artificial Intelligence for Science and Technology (QuArtist) \\ and Physics Department, Shanghai University, 200444 Shanghai, China}
	
\author{C.-Y. Pan}
\affiliation{International Center of Quantum Artificial Intelligence for Science and Technology (QuArtist) \\ and Physics Department, Shanghai University, 200444 Shanghai, China}
	
\author{L. Lamata}
\affiliation{Departamento de F\'isica At\'omica, Molecular y Nuclear, Universidad de Sevilla, 41080 Sevilla, Spain}
	
\author{E. Solano}
\email[E. Solano]{\qquad enr.solano@gmail.com}
\affiliation{International Center of Quantum Artificial Intelligence for Science and Technology (QuArtist) \\ and Physics Department, Shanghai University, 200444 Shanghai, China}
\affiliation{Department of Physical Chemistry, University of the Basque Country UPV/EHU, Apartado 644, 48080 Bilbao, Spain}
\affiliation{IKERBASQUE, Basque Foundation for Science, Plaza Euskadi 5, 48009 Bilbao, Spain}
\affiliation{Kipu Quantum, Kurwenalstrasse 1, 80804 Munich, Germany}
		
\author{F. Albarr\'an-Arriagada}
\email[F. Albarr\'an-Arriagada]{\qquad pancho.albarran@gmail.com}
\affiliation{Departamento de F\'isica, Universidad de Santiago de Chile (USACH), Avenida Ecuador 3493, 9170124, Santiago, Chile}
	
\begin{abstract}
We propose an adaptive random quantum algorithm to obtain an optimized eigensolver. Specifically, we introduce a general method to parametrize and optimize the probability density function of a random number generator, which is the core of stochastic algorithms. We follow a bio-inspired evolutionary mutation method to introduce changes in the involved matrices. Our optimization is based on two figures of merit: learning speed and learning accuracy. This method provides high fidelities for the searched eigenvectors and faster convergence on the way to quantum advantage with current noisy intermediate-scaled quantum (NISQ) computers.
\end{abstract}
	
\maketitle
	
\section{Introduction}
The emulation of biological systems has always led to disruptive bio-inspired technologies. During the last decades, machine learning (ML) emerges as an innovative technique that imitates the learning abilities of humans~\cite{Russell1995Book,Michalski2013Book,Jordan2015Science,Carleo2019}, where reinforcement learning (RL) occupies an important role. In simple terms, these protocols optimize their performance by the use of trial and error methods~\cite{Kaelbling1996JAIR,Sutton1998Book}. This class of algorithms has achieved impressive results as master players for board and video games~\cite{Silver2017Nature,Silver2018Science,Vinyals2019Nature}. On the other hand, the development of quantum computing provides a theoretical framework to break fundamental limits of classical computing~\cite{Steane1998RPP, Preskill2018Quantum, Gyongyosi2019CSR,Knill2010Nature}. With the experimental advances in quantum computing in platforms like trapped ions~\cite{Haffner2008PhysRep,Benhelm2008NatPhys,Bruzewicz2019APR}, superconducting circuits~\cite{Mariantoni2011Science,Devoret2011Science,Wendin2017RPP,Huang2020SCIS}, and photonics~\cite{Slussarenko2019RPP}, quantum supremacy was recently reached~\cite{Arute,WuPhysRevLett2021,Zhong2020Science}. Nevertheless, full-fledged fault-tolerant quantum computers are still far from reach. The development of another class of algorithms is needed.

In this manner, quantum computers were able to surpass the performance of current supercomputers for a specific task, be it quantum speckle or boson sampling. Along these lines, quantum machine learning (QML)~\cite{Schuld2015ContempPhys, Biamonte2017Nature, Dunjko2018RPP,Lamata2020MLST} is considered a natural application to surpass current classical protocols to create intelligent machines. In last years, QML has been a fruitful area, producing faster algorithms for several tasks such as linear and nonlinear algebraic problems, data classification, and variational algorithms~\cite{Harrow2009PRL, Wang2017PRA, Arrazola2019PRA, Xin2020PRA, Lloyd2020arXiv}. As in classical ML methods, also in QML the RL paradigm has received a great attention, especially for quantum control~\cite{Bukov2018PRX,Niu2019NPJ,An2019EPL,Wang2020PRL}, quantum tomography~\cite{AlbarranArriagada2018PRA, Yu2019AQT}, state preparation~\cite{Mackeprang2020QMI,Bukov2018PRB}, as well as optimization of quantum compilers~\cite{Zhang2020PRL},  among others~\cite{Dong2008IEEE,Paparo2014PRX}. This quantum computing revolution of intelligent algorithms has opened the door to develop bio-inspired quantum technologies and quantum artificial life protocols~\cite{AlvarezRodriguez2014SciRep, AlvarezRodriguez2016SciRep, AlvarezRodriguez2018SciRep, Patil2018SciRep}. In this context, random changes as mutations seem a good starting point for quantum evolutionary algorithms.

A semi-autonomous quantum eigensolver has been recently developed theoretically and experimentally~\cite{AlbarranArriagada2020MLST, Pan2020SciRep} for the calculation of eigenvectors. This algorithmic method is based on random changes on quantum states handled by single-shot measurements and feedback loops. This can be seen as a mimicking of a natural selection process, where a system evolves due to mutations (random changes) plus an abiotic environment (single-shot measurements). Since this class of algorithms employs only single-shot measurements in each feedback loop, they save a large amount of resources. To reduce the number of copies of the quantum system to be measured is indeed important, in particular when comparing to algorithms relying on expectation values such as the Variational Quantum Eigensolver (VQE)~\cite{Peruzzo2014NatCom,McClean2016NJP,Ferguson2021PRL}. In Ref.~\cite{Pan2020SciRep}, it was shown that, for a single-qubit operator, the semi-autonomous quantum eigensolver need only 200 single shot measurements, while VQE need more than 50 times more measures for similar fidelity. Random algorithms, in general, look more robust to noise if we compare with other hybrid algorithms~\cite{AlbarranArriagada2018PRA,AlbarranArriagada2020MLST}. On the other hand, random methods are designed to approach but not to match exact solution, at variance with other hybrid classical-quantum algorithms which may do that if fault-tolerant quantum computers were available. Consequently, the simplicity of semi-autonomous quantum algorithms makes them more suitable for current noisy intermediate-scale quantum (NISQ) processors, where noise is just part of the computations.

In this work, we propose a bio-inspired adaptive random quantum eigensolver (ARQE), which is strong under stochastic noise present in the gates in a quantum device. This characteristic makes our proposal suitable for NISQ devices, where the circuit depth is limited by the error of the gates among others sources. We use adaptive random mutations in the eigensolver matrices for getting high fidelities in the eigenvectors of given operators. To this end, we parametrize an arbitrary probability distribution function (PDF), where mutations are selected from. Then, we optimize with two criteria: (i) maximizing the fidelity of the learning accuracy and (ii) maximizing the learning speed via minimization of the number of iterations. The introduced ARQE algorithm is able to deliver high fidelities with faster approximations than variational methods, making it useful for approaching quantum advantage in the NISQ era.

\section{Semi-autonomous Quantum Eigensolver}

An arbitrary quantum observable is mathematically described by a Hermitian operator $\mathcal{O}$, defined by
\begin{equation}
\mathcal{O}=\sum_j \lambda_j\ketbra{\psi_j}{\psi_j},
\end{equation}
where $\lambda_j$ and $\ket{\psi_j}$ are the $j$th eigenvalue and eigenvector, respectively. A $d$-dimensional quantum system, which we will call quantum individual (QI), is characterized by its quantum state $\ket{I(\vec{\theta}_t)}$, which depends on a set of parameters $\vec{\theta}_t=(\theta_{0,t},\theta_{1,t},\dots,\theta_{n,t})$ at time $t$, with $n=2(d-1)$. The set of parameters ($\vec{\theta}_t$) can be considered as the QI's genotype. The role of the abiotic environment is given by a quantum evolution $U_E=e^{-i\mathcal{O}}$, which reads
\begin{equation}
U_E=\sum_j e^{-i\lambda}\ketbra{\psi_j}{\psi_j}.
\label{Eq02}
\end{equation}

At time $t$, we generate the QI described by $\ket{I_j(\vec{\theta}_t)}=G(\vec{\theta}_t)\ket{j}$, where $G(\vec{\theta}_t)$ is the codification gate and $\ket{j}$ is the initial state provided by the quantum processor in the computational basis, which define our $j$th solution for the eigenvectors. This codification gate plays the role of a variational ansatz, as in hybrid classical-quantum algorithms like VQE. This gate can also be decomposed in two level unitary gates as shown in Ref.~\cite{AlbarranArriagada2018PRA}. After the codification, the QI interacts with the abiotic environment, changing its state as
\begin{eqnarray}
\ket{F_j(\vec{\theta}_t)}&=&U_E\ket{I_j(\vec{\theta}_t)}=\alpha_{j,t}\ket{I_j(\vec{\theta}_t)}+\beta_{j,t}\ket{I_j^{\perp}(\vec{\theta}_t)}\nonumber\\
&=&\alpha_{j,t}\ket{I_j(\vec{\theta}_t)}+\beta_{j,t}\sum_{k\ne j}c_{k,j,t}\ket{I_k(\vec{\theta}_t)} ,
\end{eqnarray}
while satisfying the following expressions $\sum_{k\ne j}c_{k,j,t}^2=1$, $\braket{I_j^{\perp}(\vec{\theta}_t)}{I_j^{\perp}(\vec{\theta}_t)}=1$, and $\braket{I_j(\vec{\theta}_t)}{I_j^{\perp}(\vec{\theta}_t)}=0$. Now, we collapse the wave function in the basis $\{\ket{I_j(\vec{\theta}_t)}\}$ (measurement process) or, equivalently, we perform first the gate $G(\vec{\theta}_t)^{\dagger}$ and then a measurement in the computational basis $\{\ket{j}\}$. This measurement process takes the role of a death/alive (D/A) event. As the goal is to adapt the QI state to one of the eigenvectors of $\mathcal{O}$ (therefore eigenvectors of $U_E$), we consider that the QI dies if we obtain, in the measurement process, the state $\ket{m}$ with $m\ne j$. This means that if $\beta_{j,t}\ne 0$, therefore $\ket{I_j(\vec{\theta}_t)}$ cannot be eigenvector of $\mathcal{O}$. In the other case, if we measure the state $\ket{j}$, then the QI survives to the D/A event, and it is a candidate for eigenvector. In the following iteration, $t+1$, we create the QI described by $\ket{I_j(\vec{\theta}_{t+1})}=G(\vec{\theta}_{t+1})\ket{j}$, where we define
\begin{eqnarray}
&&\vec{\theta}_{t+1}=(\theta_{0,t+1},\theta_{1,t+1},\dots,\theta_{n,t+1}) , \nonumber\\
&&\theta_{k,t+1}=\theta_{k,t}+\pi \epsilon_{k,t}\cdot(1-\delta_{m,j}) .
\label{anglechange}
\end{eqnarray}
Here, $m$ is the measurement outcome of the previous iteration $t$, $\delta_{m,j}=1\iff m=j$ and  $\delta_{m,j}=0$ for $m\ne j$, while $\epsilon_{k,t}$ is a random number in the range $[-1,1]$ with a PDF $\mathcal{D}_{t+1}$.

\begin{figure}[t]
\centering
\includegraphics[width=0.8\linewidth]{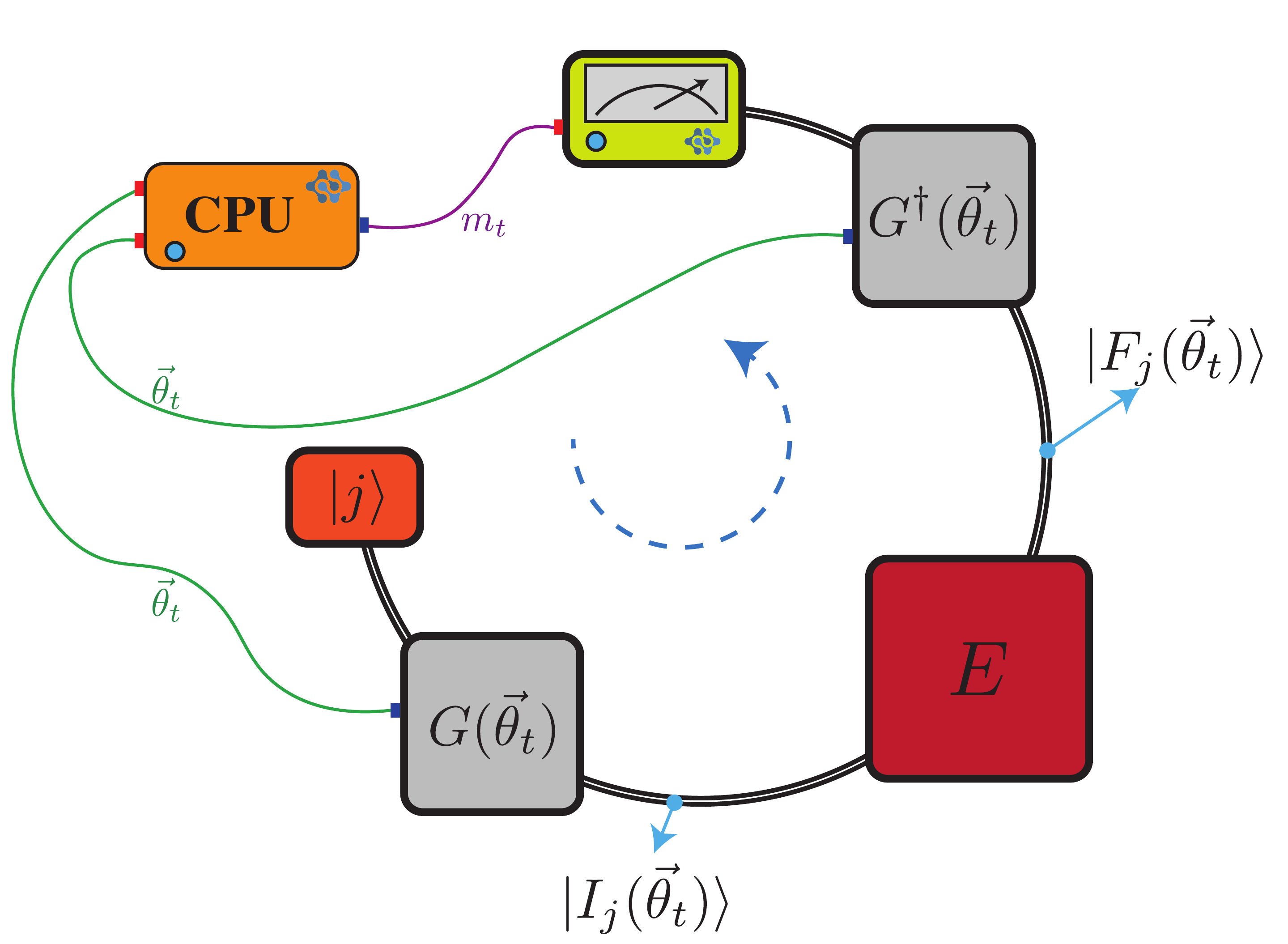}
\caption{Scheme of a bio-inspired adaptive random quantum algorithm. The individual is mapped on the encoding gate $G(\vec{\theta}_t)$, the abiotic environment is represented by gate $E$, and the dead-alive probability is given by the decoding gate  $G^{\dagger}(\vec{\theta}_t)$, and a measurement in the computational basis $\{\ket{j}\}$. We introduce changes in the encoding and decoding matrix (mutations) using a classical feedback loop (green lines) which depends on the classical communication of the measurement outcome (purple line).}
\label{Fig01}
\end{figure}

Equation (\ref{anglechange}) introduces mutations in the genotype only if the QI dies ($m\ne j$), which means that we create a new QI for time $t+1$. Moreover, if the QI survives, we replicate the same QI for time $t+1$. Additionally, we change the PDF in each step according to a suitable reward/punishment (R/P) criterion~\cite{AlbarranArriagada2020MLST,Pan2020SciRep}. In general, the latter will increase the probability to obtain stronger mutations (major changes) each time that the local goal is not reached (dead), and decrease the probability to obtain stronger mutations (minor changes) each time that we reach the local goal (alive). We have several R/P criteria to modify the PDF in time to ensure compliance of the previous requirement, which will be specified later. In next section, we will describe a general parametrization of a symmetric PDF suitable to be optimized and obtain a correct R/P criterion to find the eigenvectors of $\mathcal{O}$. Figure \ref{Fig01} shows a scheme of the adaptive algorithm.

We can summarize our protocol as follows for the $t$th iteration:
\begin{enumerate}
\item Initialize our quantum device in the state $\ket{j}$. In general, for quantum computers it is usually the state $\ket{0}$.
\item Apply the codification gate $G(\vec{\theta}_t)$ with parameters $\vec{\theta}_t$. As mentioned above, it can be decomposed in two-level unitary operations.
\item Apply the unitary evolution $U_E$ given by the environment.
\item Apply the gate $G^{\dagger}(\vec{\theta}_t)$ to decode the state.
\item Measure the resulting state which is given by
\begin{equation}
\ket{\Psi_{j,t}}=G^{\dagger}(\vec{\theta}_t)U_EG(\vec{\theta}_t)\ket{j}.
\end{equation}
in the computational basis.
\item Update the parameters  $\vec{\theta}_t$ for the iteration $t+1$ according to the measurement output as is given in Eq.~(\ref{anglechange}).
\item Repeat the process.
\end{enumerate}

We point out that random algorithms provide a fast approximation for the eigenspectrum of quantum observables. The total or partial knowledge of the eigenspectrum of an unknown operator is a crucial task for efficient classification of quantum states or to boost quantum optimizers such as the recent proposed algorithms based on Digitized-Counterdiabatic Quantum Computing (DCQC)~\cite{HegadePhysRevApp2021}. Also, for semi-autonomous quantum devices the capability to adapt a quantum state into an eigenstate could help to develop more sophisticated machines (see Ref.~\cite{AlbarranArriagada2020MLST}). Therefore, the enhancement of such algorithms is worth studying.

\section{Optimization algorithm}
From Eq.~(\ref{anglechange}), we can observe that the core of the algorithm is the random change given by the random variable $\epsilon_{k,t}$. Then, to optimize the method described in the previous section, we need optimize the PDF that defines $\epsilon_{k,t}$. To do this, we parametrize the probability cumulative function (PCF) of a random number generator by means of the inverse transform sampling technique (ITST). According to the ITST we can generate a random variable $X$ in the range $[-\infty,\infty]$ with PCF $F_X(x)$, by the use of another random variable $Y$ in the range $[0,1]$ with uniform PDF ($D_Y(y)=1$). We know that the probability for the values of $X$ to be in the range $[a,b]$ is
\begin{equation}
P(a<X<b)=\int_{a}^bD_X(x)dx,
\end{equation} 
and the relation between the PDF ($D_X$) and the PCF ($F_X$) is
\begin{equation}
F_X(x)=\int_{-\infty}^xD_X(\bar{x})d\bar{x}=P(-\infty<X<x).
\end{equation} 
Finally, using the ITST, we have that the random variable $X$ with PDF $D_X(x)$ is given by $X=F_X^{-1}(Y)$. Therefore, by the parametrization of the PCF $F_X$, we are parametrizing the random number generator.

\subsection{Parametrization of $F_X$}

As $F_X(x)$ is a PCF of a random variable $X$, it is a monotonically increasing function, with $F_X(-\infty)=0$, and $F_X(\infty)=1$. Moreover, as the random variable represents a mutation in our algorithm, the PDF needs to be symmetric. Therefore, we impose the extra condition over the PCF,
\begin{equation}
F_X(x)=1-F_X(-x),
\end{equation}
which implies $F_X(0)=\frac{1}{2}$. Finally, as we consider mutations, we will focus in the generation of a random variable in the range $[-1,1]$, which means $F_X(-1)=0$, and $F_X(1)=1$.

\begin{figure}[t]
\centering
\includegraphics[width=0.9\linewidth]{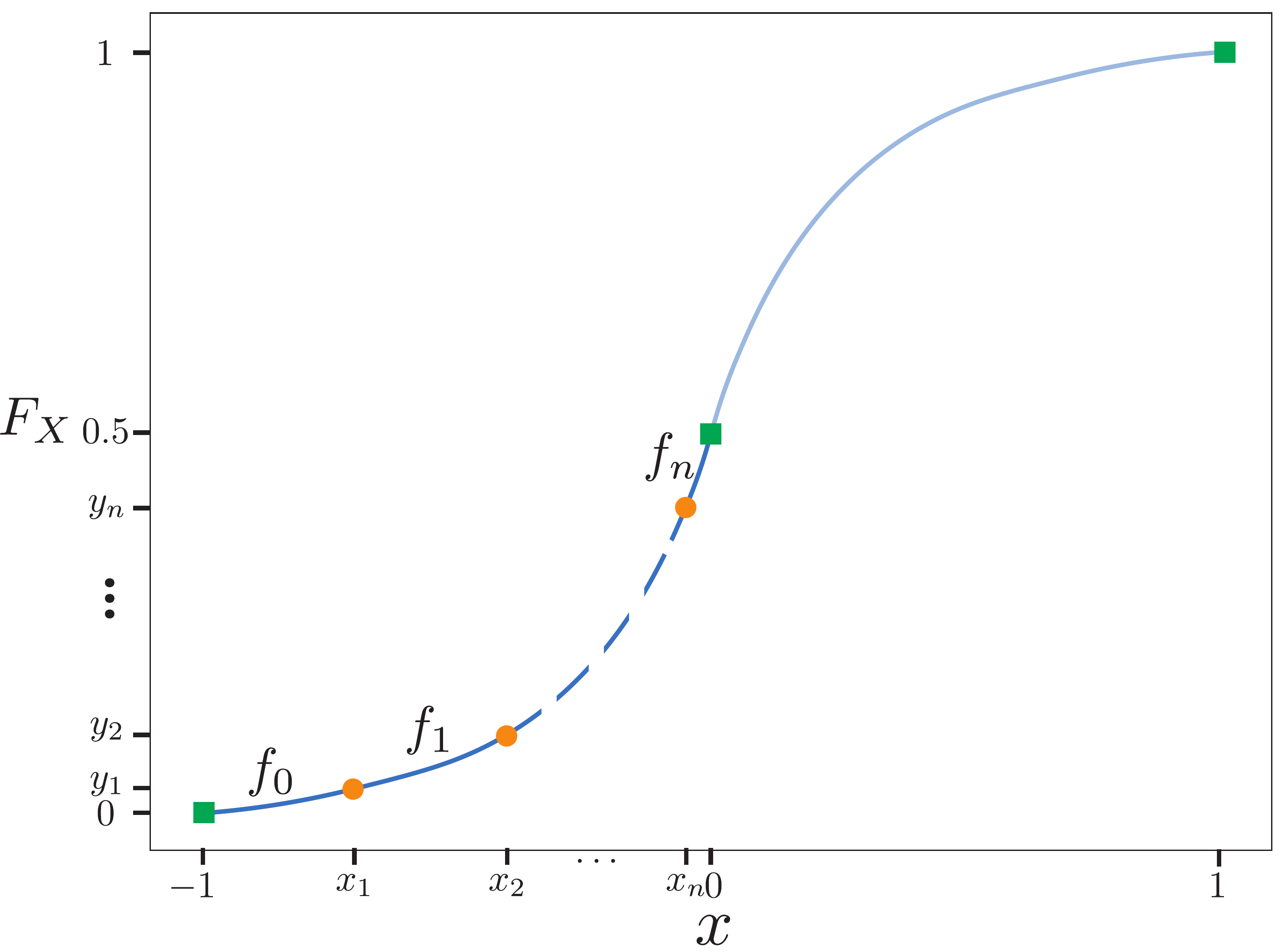}
\caption{Points $(x_j,y_j)$ (orange circles), and the monotonic and symmetric interpolation defined by Eq. (\ref{Eq08}). The green squares represent the points $p_0=(-1,0)$, $p_{n+1}=(0,0.5)$, and $(1,1)$, which are fixed to ensure that the interpolation corresponds to a valid symmetric PCF. $f_j$ represents the function by parts that interpolates the points $p_{j}$ and $p_{j+1}$.}
\label{Fig02}
\end{figure}

To parametrize the PCF, we consider the following vectors $\vec{x}=\{x_0=-1,x_1,...,x_n,x_{n+1}=0\}$ and $\vec{y}=\{y_0=0,y_1,...,y_n,y_{n+1}=0.5\}$ in ascending order. These two vectors define the points $\mathcal{P}=\{p_j=(x_j,y_j)\}$. Now, by considering a monotonic interpolation method through the points in the set $\mathcal{P}$, we can obtain a parametrized function $\mathfrak{F}(x,\vec{x},\vec{y})$, depending on $2n$ parameters (the end points are fix). Using this, we can construct a parametrized PCF $F_X(x,\vec{x},\vec{y})$ as (see Fig. \ref{Fig02}) 
\begin{eqnarray}
&&F_X(x,\vec{x},\vec{y})=\mathfrak{F}(x,\vec{x},\vec{y}),\quad x<0 , \nonumber\\
&&F_X(x,\vec{x},\vec{y})=1-\mathfrak{F}(x,\vec{x},\vec{y}),\quad x>0 .
\label{Eq08}
\end{eqnarray}
Here (see Ref.~~\cite{Steffen1990AstroA}),
\begin{eqnarray}
&& \mathfrak{F}(x,\vec{x},\vec{y})=f_j(x) , \,\,\,  x\in[x_j,x_{j+1}] , \nonumber \\
&& f_j(x)=a_j(x_j-x)^3+b_j(x_j-x)^2+c_j(x_j-x)+d_j , \nonumber \\
\end{eqnarray}
and
\begin{eqnarray}
&&a_j=\frac{y_j'+y'_{j+1}-2s_j}{h_j^2},\quad b_j=\frac{3s_j-2y_j'-y_{j+1}'}{h_j},\nonumber\\
&&c_j=y_j', \qquad d_j=y_j,
\end{eqnarray}
where
\begin{equation}
s_j=\frac{y_{j+1}+y_{j}}{x_{j+1}+x_{j}},\quad h_j=x_{j+1}+x_{j},
\end{equation}
and
\begin{equation}
y_j'=\frac{d}{dx}f_j|_{x=x_j}=\frac{d}{dx}f_{j-1}|_{x=x_j}.
\end{equation}
Now, we approximate the derivative of the function $f_j$ by
\begin{eqnarray}
&& y_j'=0,\quad \textrm{if}\; (s_{j_1}\cdot s_j)\le 0\nonumber\\
&& y_j'=2\cdot \textrm{sign}(s_j)\cdot \abs{s}^{\textrm{min}}_{j-1,j} , \nonumber \\ && \quad \textrm{if}\; \abs{p_{j}}>2\abs{s}^{\textrm{min}}_{j-1,j} 
y_j'=p_j\quad \textrm{otherwise} ,
\label{Eq13}
\end{eqnarray}
where
\begin{eqnarray}
&& p_j = \frac{s_{j-1}h_{j}+s_{j}h_{j-1}}{h_{j-1}+h_{j}}, \nonumber \\
&& \abs{s}_{j-1,j}^{\textrm{min}} = \textrm{min}(\abs{s_{j-1},\abs_{s_j}}).
\end{eqnarray}

We note that, according to Eq. (\ref{Eq13}), we can estimate $y'_j$ only for $j\in\{1,...,n\}$. We impose the border condition $y'_0=0$, and the symmetry condition $y_{n+1}'=s_{j-1}$.

Finally, we introduce the R/P criteria that we will use in the rest of the article for the ARQE protocol. As we need to change the PDF of the random number generator, we will define the PCF as in Eq. (\ref{Eq08}) but with parameters that will change in each iteration, namely,
\begin{equation}
F_X=F_X(x,\vec{x}_k,\vec{y}_k) .
\end{equation}
Here, $\vec{x}_k,\vec{y}_k$ are defined for $k$th iteration of our algorithm as
\begin{equation}
\vec{x}_k=w_k\cdot \vec{x}_{k-1}\quad \vec{y}_k=w_k\cdot \vec{y}_{k-1} ,
\end{equation}
with
\begin{equation}
w_k=[p+(r-p)\delta_{m,j}]w_{k-1} ,
\end{equation}
where $r<1$ is the reward constant, $p>1$ the punishment constant, $m$ is the measured outcome, and $j$ the desired outcome defined in previous section. Also, we require for convergence purposes that $1\le r\cdot p$. We define that the algorithm converges after $N$ iterations if $w_N<\Delta$, where $\Delta$ is the tolerance of our algorithm.

\subsection{Optimization of $F_X$}

As the proposed ARQE method achieves the result in a stochastic way, we optimize the random number generator using two criteria. In the first one, we define the cost function as the mean number of iterations needed for convergence ($\bar{N}$), obtaining the values of $\vec{x}_0$ and $\vec{y}_{0}$ which minimize $\bar{N}$. For the second one, we define the fidelity after $\ell$ iterations as 
\begin{equation}
\mathcal{F}_{\ell}=\abs{\bra{\psi_j}G(\vec{\theta}_\ell)\ket{j}}^2.
\end{equation}
In this case, we define the mean fidelity after $\ell$ iterations ($\bar{\mathcal{F}}_{\ell}$) as the cost function, obtaining the values of $\vec{x}_{opt}$ and $\vec{y}_{opt}$ which maximize $\bar{\mathcal{F}}_{\ell}$. For the calculation of the mean values we consider $1000\cdot n_q$ independent repetitions of the algorithm, where $n_q$ is the number of qubits involved in the algorithm. We remark that this optimization process depends on the quantum operator to be diagonalized. In order to carry out a general optimization, we consider the optimization of different operators (100 cases), which means finding $\vec{x}_{opt}$ and $\vec{y}_{opt}$ for a set of different cases, which could be used for the prediction of $\vec{x}_{opt}$ and $\vec{y}_{opt}$ for new operators.

We note that for the learning accuracy, the estimation of the fidelity $\mathcal{F}_{\ell}$ requires a tomography process and also the previous knowledge of the eigenstates $\ket{\psi_j}$, which means that is impractical from an experimental point of view. Nevertheless, this is interesting to analyse from a pedagogical point of view. Also, the learning accuracy strategy can be enhanced by the numerical simulations and extrapolated for complex systems to avoid experimental limitations. 
\begin{figure}[t]
\centering
\includegraphics[width=1\linewidth]{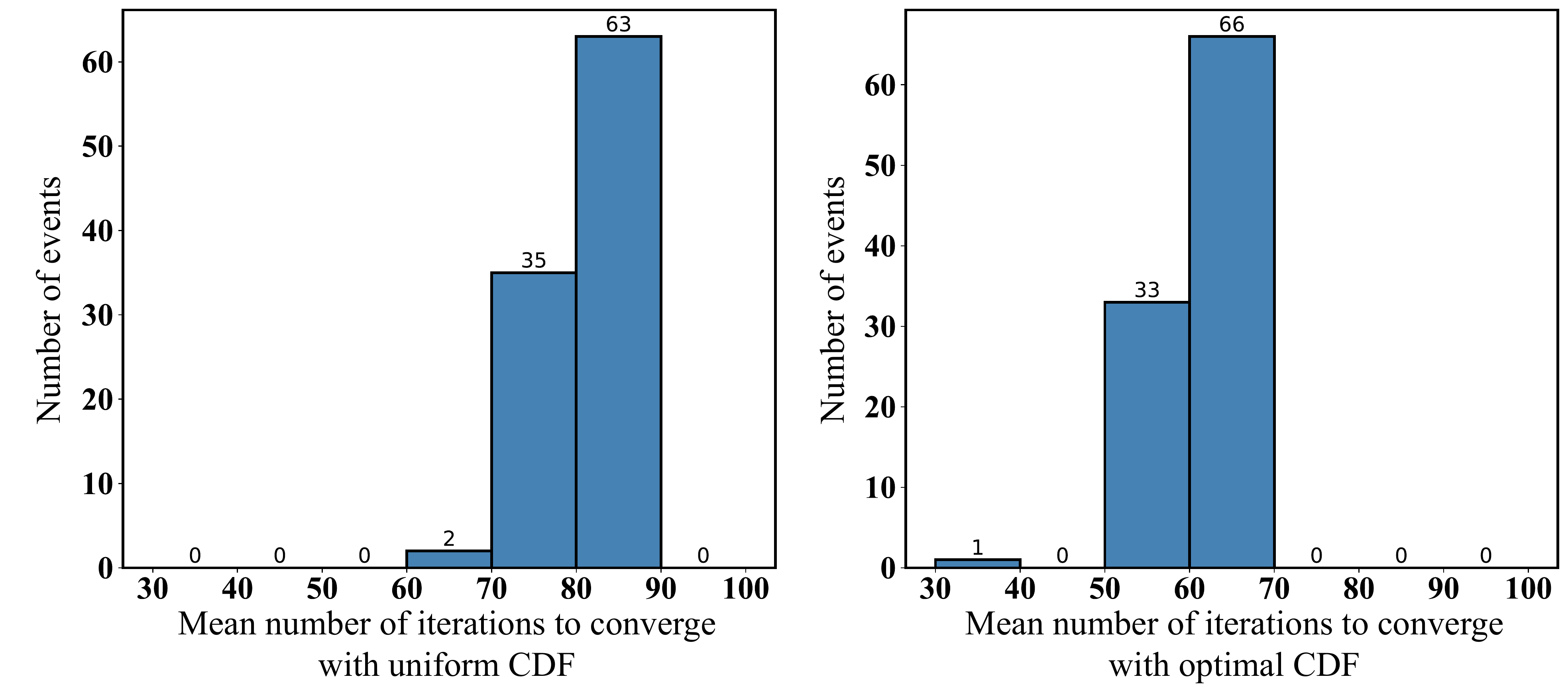}
\caption{Histogram for mean number of interactions to convege without (left panel) and with optimization (right panel) for 100 different $U(\theta,\phi,\lambda)$. Both panels are for optimization of learning speed.}
\label{Fig03}
\end{figure}

\section{Results}

We consider single-qubit operators of the form
\begin{equation}
\mathcal{O}(a_I,a_x,a_y,a_z)=a_I\mathbb{I}+a_x\sigma_x+a_y\sigma_y+a_z\sigma_z.
\end{equation}
In this case, the codification matrix is a general single-qubit unitary matrix given by
\begin{equation}
U(\theta,\phi,\lambda)=\begin{pmatrix}
\cos(\theta/2) & -e^{i\phi}\sin(\theta/2)\\
e^{i\lambda}\sin(\theta/2) & e^{i(\phi+\lambda)}\cos(\theta/2) 
\end{pmatrix} ,
\end{equation}
which depends on three parameters (genes). We use 100 different sets of genes chosen randomly in the range $[0,2\pi]$ to cover different situations. We consider $\vec{x}=[-1, x_1, x_2, 0]$ and $\vec{y}=[0, y_1, y_2, 0.5]$, where we define $X_{in}=[x_1, x_2]$ and $Y_{in}=[y_1, y_2]$ for the PCF parametrization and optimization. Here, we optimize the iteration number needed until the convergence parameter, $w_N$, surpasses a threshold $\Delta=0.9$ (learning speed). The 100 results of the different optimizations are summarised in Fig. \ref{Fig03}. From this figure, we can see that the mean-iteration number decreases, from the data set of this case (see table \ref{table1} in the appendix), we have that the mean value of the mean-iteration number for the optimized case is $\bar{N}\approx61$, while for the case without optimization is $\bar{N}\approx82$ which means a reduction of 25.4$\%$. The cases without optimization refers to a uniform PDF for the mutation process. We also need to mention that in this case the fidelity of the obtained solution remains almost constant (see Fig.~9 in the appendix), obtaining less iterations to almost converge to the same solution.

\begin{figure}[b]
\centering
\includegraphics[width=1\linewidth]{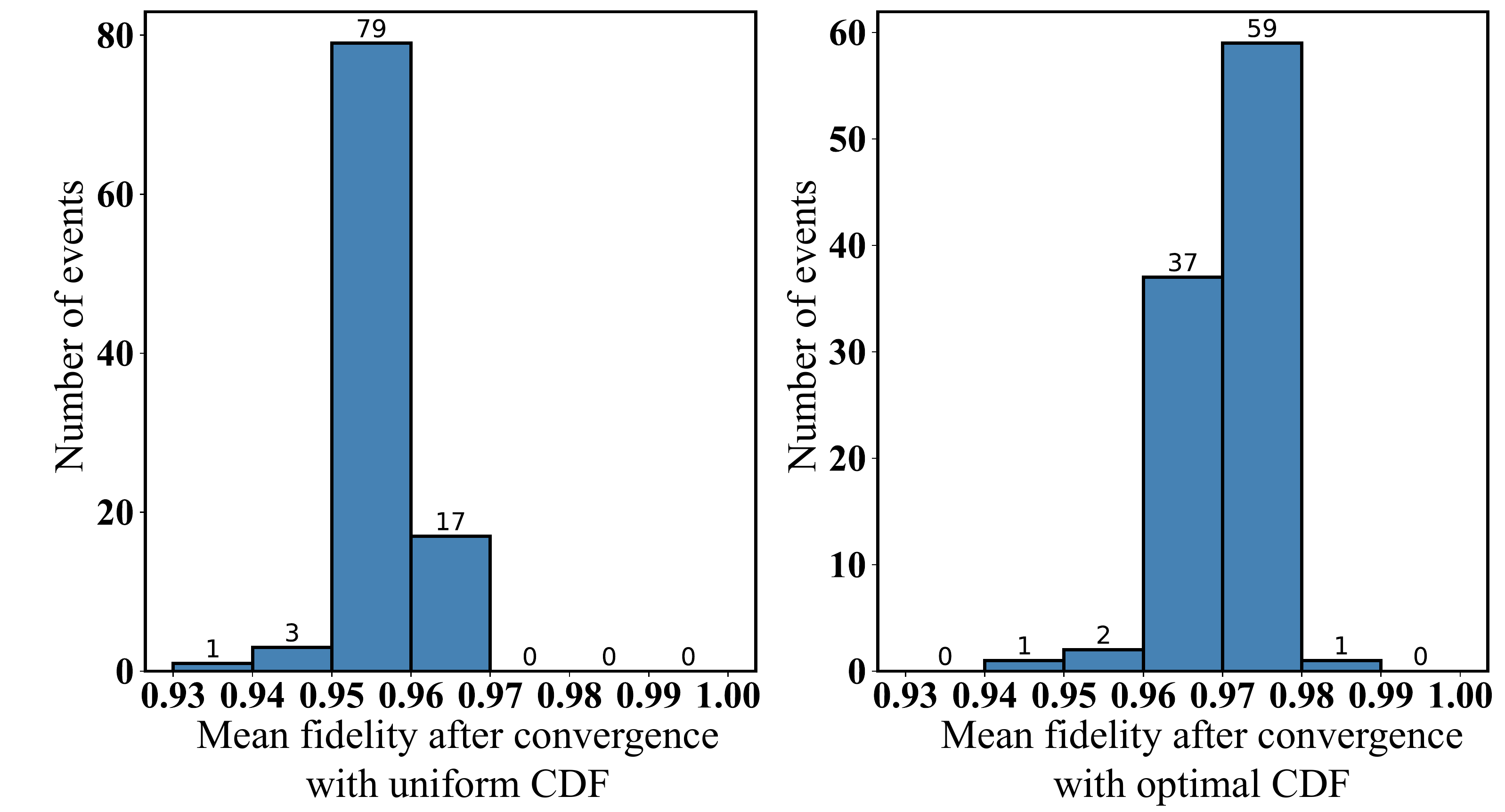}
\caption{Histogram for mean fidelity after convergence without (left panel) and with optimization (right panel) for 100 different $U(\theta,\phi,\lambda)$, for optimization of learning accuracy.}
\label{Fig07}
\end{figure}

In addition, we also perform the optimization fixing the number of iterations $N=80$ and minimizing the convergence parameter, which implies maximization of the fidelity (learning accuracy). We choose again 100 random unitary operators $U(\theta,\phi,\lambda)$ for the environment. Figure \ref{Fig07} summarises the data for the mean fidelity with and without optimization for this case, which shows a clear increase of the fidelity. From the data set of this case (see table \ref{table2} in the appendix), we have that the mean value of the fidelity increases from $\bar{F}\approx0.95$ without optimization to $\bar{F}\approx0.97$ with optimization, increasing the learning accuracy of our protocol.

\begin{figure}[t]
\centering
\includegraphics[width=1.0\linewidth]{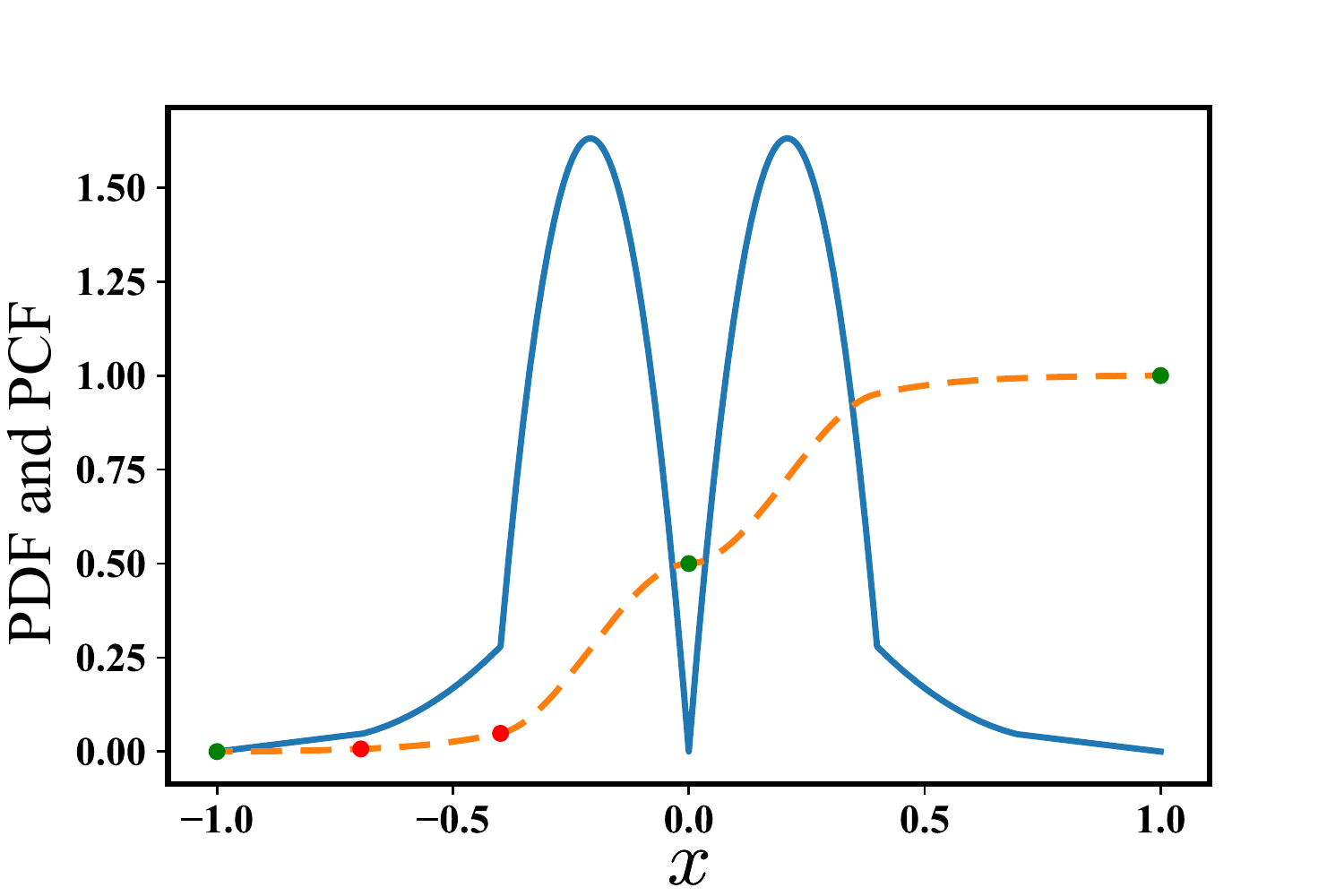}
\caption{Optimal PDF (solid blue line) and the PCF (dashed orange line) for learning speed for $\tau\hat{\mathcal{O}}=\sigma_x$. Red dots are the points related to the parametrization, green dots are the fix points of our PCF.}
\label{Fig08}
\end{figure}
\begin{figure}[b]
\centering
\includegraphics[width=1.0\linewidth]{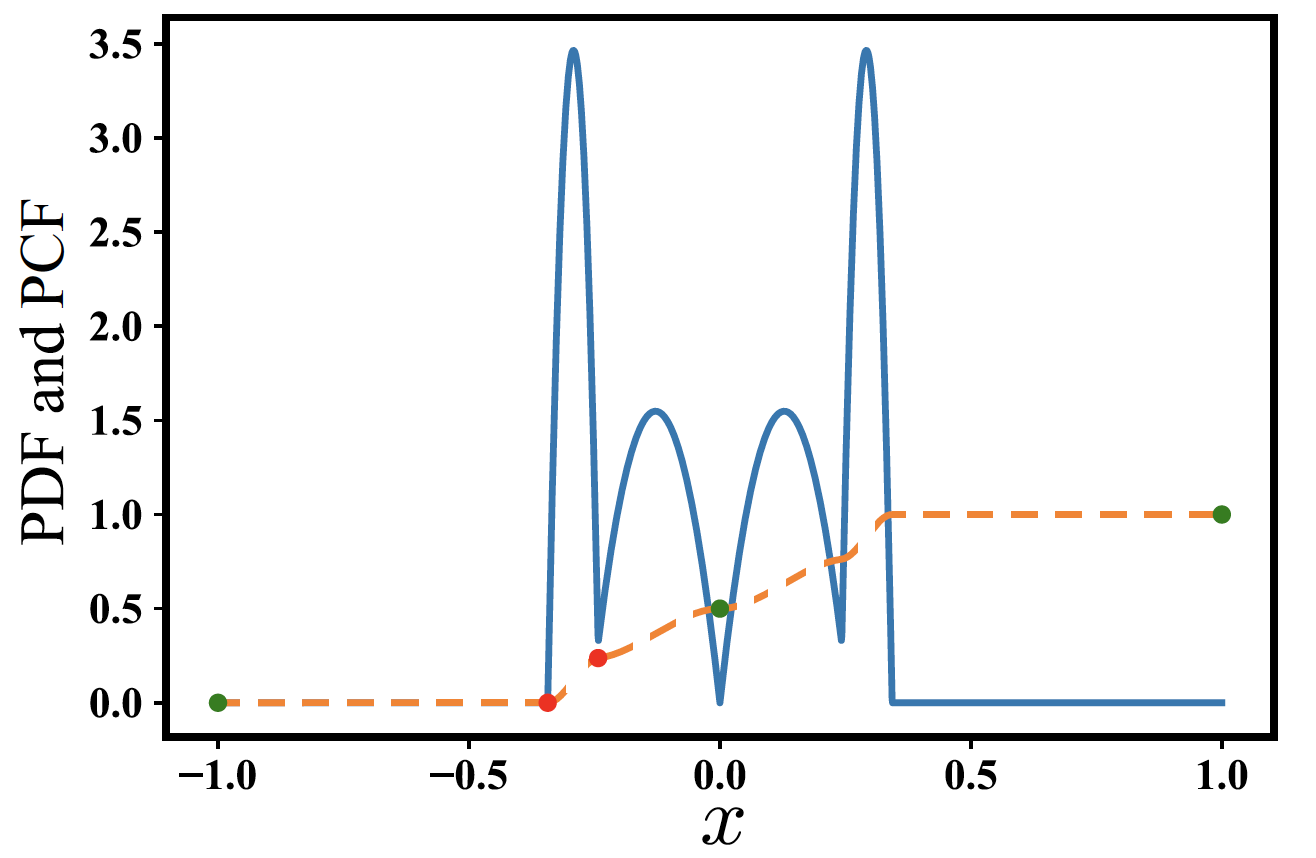}
\caption{Optimal PDF (solid blue line) and the PCF (dashed orange line) for learning accuracy for $\tau\hat{\mathcal{O}}=\sigma_x$. Red dots are the points related to the parametrization, green dots are the fix points of our PCF.}
\label{Fig09}
\end{figure}

Figure \ref{Fig08} shows an example for the optimal PCF and its corresponding PDF for the optimization of the learning speed. Specifically, the genes are $\theta=2,\,\phi=\frac{\pi}{2},\,\lambda=\pi$, which correspond to $\tau\mathcal{O}=\sigma_x$. The optimal parameters are $x_1=-0.69513535,\,x_2= -0.3989989,\,y_1=0.00706757,\,y_2=0.04842301$. We can see that the optimal PDF has two symmetric peaks, which means that the optimal adaptation appears when the most probable mutation is different from zero and approaches zero when the quantum individual becomes adapted. On the other hand, Fig. \ref{Fig09} shows the optimal PCF and PDF using the same genes but for the optimization of the learning accuracy. Here, the optimal parameters are $x_1=-0.34315537,\,x_2=-0.24266087,\,y_1=0,\,y_2=0.23737731$. 

\begin{figure}[h]
\centering
\includegraphics[width=1.0\linewidth]{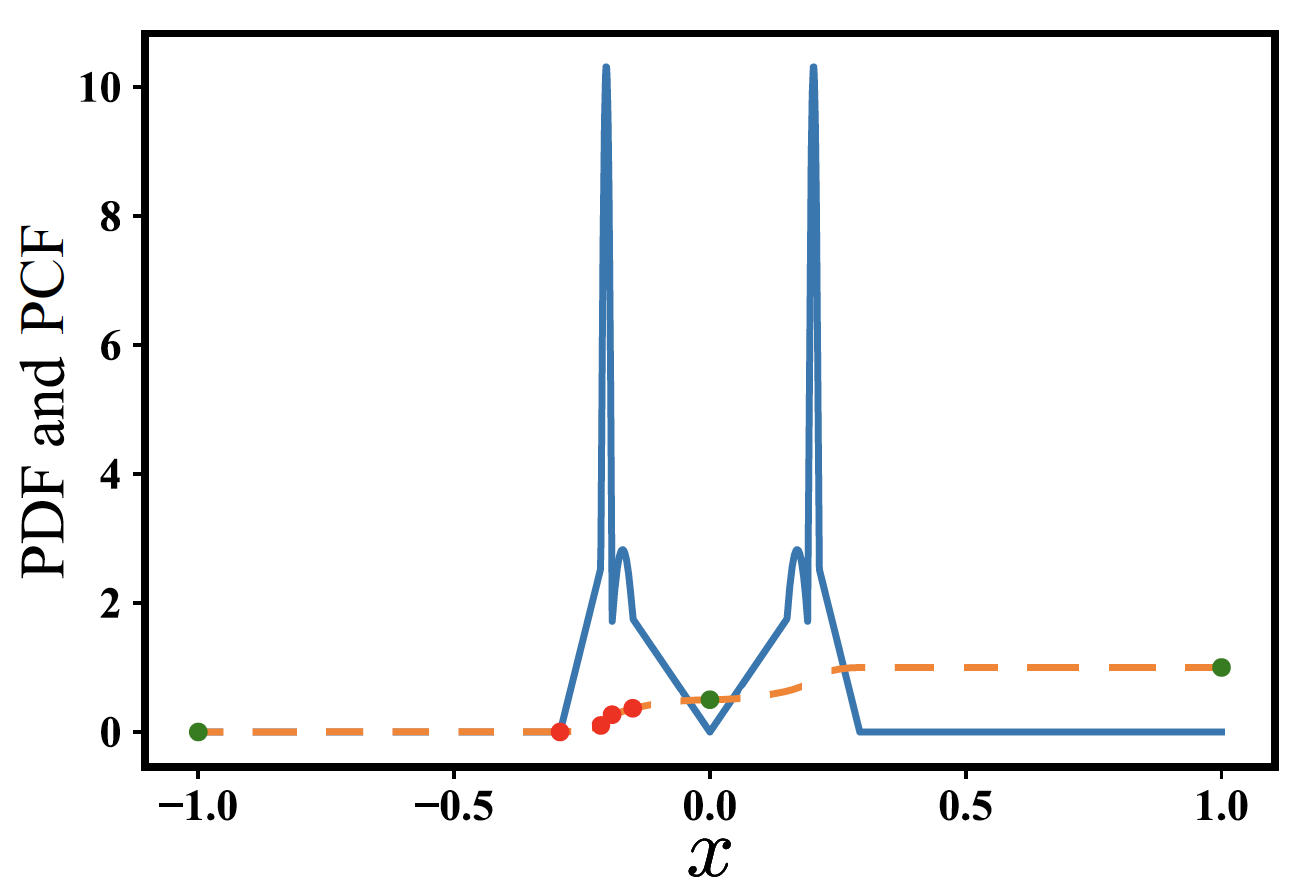}
\caption{Optimal PDF (blue solid line) and the PCF (orange dashed line) for learning speed for $\tau\mathcal{O}$ given by Eq (\ref{Eq21}). Red dots are the points related to the parametrization, green dots are the fix points of our PCF.}
\label{Fig11}
\end{figure}

\begin{figure}[h]
\centering
\includegraphics[width=1\linewidth]{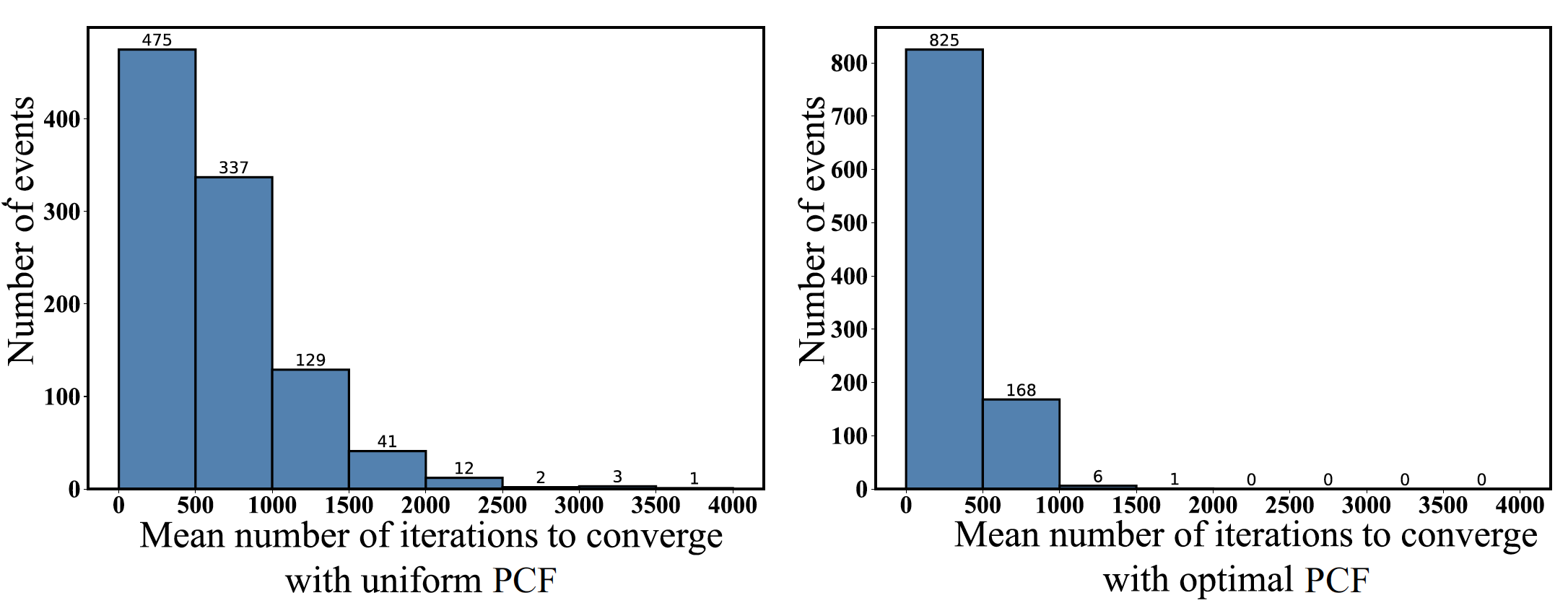}
\caption{The histogram comparison of the mean number of iterations to converge without (left panel) and with optimization (right panel) for Eq (\ref{Eq21}).}
\label{Fig12}
\end{figure}

Finally, we present two two-qubit examples. For the first one, we consider a non-degenerate operator given by
 \begin{equation}
\quad\tau\mathcal{O}=\begin{pmatrix}
		\pi & -\frac{\pi}{2} & -\frac{\pi}{4} & -\frac{\pi}{4} \\
		-\frac{\pi}{2} & \pi & -\frac{\pi}{4} & -\frac{\pi}{4} \\
		-\frac{\pi}{4} & -\frac{\pi}{4} & \frac{\pi}{2} & 0\\
		-\frac{\pi}{4} & -\frac{\pi}{4} & 0 & \frac{\pi}{2} 
	\end{pmatrix},
	\label{Eq21}
\end{equation}
with the following eigenvectors and eigenvalues
\begin{eqnarray}
	\ket{\mathcal{E}^{(0)}}=&&\frac{1}{2}(\ket{00}+\ket{01}+\ket{10}+\ket{11}), \quad \alpha^{(0)}=0, \nonumber\\
	\ket{\mathcal{E}^{(1)}}=&&\frac{1}{\sqrt{2}}(\ket{10} - \ket{11}), \quad \alpha^{(1)}=\frac{\pi}{2}, \nonumber\\
	\ket{\mathcal{E}^{(2)}}=&&\frac{1}{2}(\ket{00}+\ket{01}-\ket{10}-\ket{11}), \quad \alpha^{(2)}=\pi ,\nonumber\\
	\ket{\mathcal{E}^{(3)}}=&&\frac{1}{\sqrt{2}}(\ket{00}-\ket{01}), \quad \alpha^{(3)}=\frac{3\pi}{2}.
	\label{Eq22}
\end{eqnarray}
As the eigenvalues of this operator are equidistant, then the ARQE needs a large number of iterations to converge. It is due to the fact that the unitary evolution in Eq.~(\ref{Eq02}) is sensitive to the gap between the eigenvalues, reaching the eigenvectors with large gap easier than the closer one, accelerating our algorithm as is shown in Ref.~\cite{Pan2020SciRep} via numerical inspection. In this case, we use a four points parametrization, which means that $X_{in}=[x_1,x_2,x_3,x_4]$ and $Y_{in}=[y_1,y_2,y_3,y_4]$.
The corresponding PDF and PCF for optimal learning speed is shown in Fig.~\ref{Fig11}, while the optimal parameters are $x_1=-0.29285222,\,x_2=-0.2132477,\,x_3=-0.19124688,\,x_4=-0.15079198,\,y_1=2.85749338e-06,\,y_2=0.101155582,\,y_3=0.268075636,\,y_4=0.367572655$. The mean value of the mean-iteration number for the optimized case is $\bar{N}\approx355$ while for the case without optimization is $\bar{N}\approx654$ which means a reduction of 45.7$\%$.  The data is summarized in the histogram of Fig~\ref{Fig12}. The mean fidelities for four eigenstates without optimization are $F_0=0.952,F_1=0.947,F_2=0.941$ and $F_3=0.938$. The mean fidelities with optimization are $F_0=0.946,F_1=0.941,F_2=0.933$ and $F_3=0.930$. Therefore, we do not obtain appreciable changes in the fidelity but a considerable reduction of the iterations.

An interesting result is that we can see from Fig.~\ref{Fig08}, Fig~\ref{Fig09} and Fig.~\ref{Fig11} that the optimal PDF shows peaks in the mutation probability far from zero, which approaches zero when the QI starts to be adapted. It means that in the optimal mutations process, the changes close to zero have almost null probability, favoring the mutations in discrete regions of values for the random variable.

The second example is the molecular hydrogen Hamiltonian with a bond length of 0.2 [\AA]. In this case, the environment is given by
\begin{equation}
\tau\mathcal{O}=g_0\mathbb{I}+g_1Z_0+g_2Z_1+g_3Z_0Z_1+g_4Y_0Y_1+g_5X_0X_1 ,
\end{equation}
with $g_0 = 2.8489$, $g_1 = 0.5678$, $g_2 =-1.4508$, $g_3=0.6799$, $g_4=0.0791$, $g_5=0.0791$. The eigenvectors for this case are
\begin{eqnarray}
	\ket{\mathcal{E}^{(0)}}=&&-0.03909568\ket{01}+0.99923547\ket{10},\quad \nonumber\\
	\ket{\mathcal{E}^{(1)}}=&&\ket{00},\nonumber\\
	\ket{\mathcal{E}^{(2)}}=&&0.99923547\ket{01}+0.03909568\ket{10}, \quad \nonumber\\
	\ket{\mathcal{E}^{(3)}}=&&\ket{11}, \quad 
\end{eqnarray}
and the eigenvalues
\begin{eqnarray}
&&\alpha^{(0)}=0.14421033,\, \quad \alpha^{(1)}=2.6458 , \nonumber\\ 
&&\alpha^{(2)}=4.19378967,\,\alpha^{(3)}=4.4118 ,
\end{eqnarray}
respectively. If we choose $r=0.9$ and the convergence condition is $w< 0.1$, we need at least 21 single-shot measurements, while for three eigenvectors (the fourth one if orthogonal to the others) we need at least 63 single-shot measurements. However, in this case, the mean iteration is 65 without optimization, which means that the uniform distribution is the optimal one.

We need to mention that our ARQE algorithm is sensitive to the number of different eigenvalues; therefore, for a degenerate case, the algorithm will be faster, but as the degenerate space defines an eigensubspace, instead of a set of eigenvectors, the result is not unique.

Finally, we highlight that in this work, we do not focus in the implementation of the evolution $U_E$, which in many platforms can be nontrivial. Our ARQE algorithm mainly focus on the efficient extraction, with respect to the number of single-shot measurements, of relevant information (eigenvectors) from a quantum evolution. Moreover, our proposal is not limited to digital NISQ computers, it can be used in analog or digital-analog quantum paradigms in order to implement the evolution $U_E$ in a more natural way.

\section{Conclusions}
We have developed a random optimization protocol for proposing bio-inspired ARQE algorithms based on a PCF parametrization defining a random number generator. The latter is responsible for the mutation process, allowing the quantum individual to adapt, and is at the core of this class of algorithms. We develop these ARQE methods according to two different criteria, learning speed and learning accuracy. In this sense, this work contributes to the search of efficient strategies for random algorithms, providing good approximate solutions with fewer resources concerning other random algorithms~\cite{AlbarranArriagada2018PRA,Yu2019AQT,AlbarranArriagada2020MLST,Pan2020SciRep}. These have shown, in turn, improvements in the number of single-shot measurements when compared to hybrid classical-quantum algorithms~\cite{Pan2020SciRep}. 

Moreover, our algorithm focuses on the eigenvector of an operator whose eigendecomposition is unknown. This can be useful in the characterization of physical interactions, as well as for fast approximations in optimization algorithms reducing the searching space and therefore speeding up the minimization of Hamiltonians. In addition, the ARQE algorithm finds the eigenvectors independently of their eigenenergy, being suitable for the estimation of high energy orbitals in quantum chemistry. On the other hand, as we consider stochastic algorithms, the scalability of our proposal requires a deep study of statistical mechanics, which is out of the scope of the present work. 
Finally, we need to highlight that the goal of this work is to provide an easy formulation to parametrize a random number generator, which is the core of random algorithms like in Ref. \cite{AlbarranArriagada2018PRA,Yu2019AQT,AlbarranArriagada2020MLST,Pan2020SciRep}. This parametrization allows one to optimize the PDF of the random variable, enhancing the performance of the mentioned protocols, as is shown by the numerical results.

We expect that this kind of effort contributes to approaching quantum advantage in available or improved NISQ devices. It is noteworthy to mention that ARQE protocols may also be used as pre-processing for sophisticated algorithms, including VQE, quantum phase estimation methods or DCQC.

\section*{Acknowledgments}
We acknowledge financial support from Junta de Andaluc\'ia (P20-00617 and US-1380840), Shanghai STCSM (Grant No. 2019SHZDZX01-ZX04) and ANID Subvención a la Instalación en la Academia SA77210018.

\clearpage
\onecolumngrid
\appendix

\section{Learning speed data}

The next table collect all the data for the 100 different instances for the choose of $\{\theta,\phi,\lambda\}$ and their  $X_in$ and $Y_in$ for the optimal PCF. Also the table shows the number of iterations needed for convergence using an uniform PDF ($N$) and the optimal one ($N_{opt}$), as well as the fidelity for the optimal ($F_{opt}$) and non-optimal PDF ($F$). In this case we are optimizing the learning speed. 

\begin{longtable*}{|l|l|l|l|l|l|l|l|l|l|}
\hline
\textbf{} & \cellcolor[HTML]{BDC0BF}{\color[HTML]{000000} \bm{$\theta$}} & \cellcolor[HTML]{BDC0BF}\bm{$\phi$} & \cellcolor[HTML]{BDC0BF}\bm{$\lambda$} & \cellcolor[HTML]{BDC0BF}$X_{in}$ & \cellcolor[HTML]{BDC0BF}$Y_{in}$ & \cellcolor[HTML]{BDC0BF}$N$ & \cellcolor[HTML]{BDC0BF}$N_{opt}$ & \cellcolor[HTML]{BDC0BF}$F$ & \cellcolor[HTML]{BDC0BF}$F_{opt}$ \\ \hline
\cellcolor[HTML]{DBDBDB}\textbf{1} & 2.83651 & 2.51483 & 5.79311 & -0.352607301, -9.4996[-05] & 1.65031947[-21], 0.499956565 & 84 & 61 & 0.98860 & 0.98812 \\ \hline
\cellcolor[HTML]{DBDBDB}\textbf{2} & 2.60239 & 2.91385 & 1.94757 & -0.36580222, -0.25487967 & 0.01084567, 0.18360378 & 83 & 61 & 0.98875 & 0.98615 \\ \hline
\cellcolor[HTML]{DBDBDB}\textbf{3} & 1.62294 & 2.66070 & 0.16163 & -0.64338053, -0.46788155 & 0.18541396, 0.19697403 & 75 & 63 & 0.98722 & 0.98714 \\ \hline
\cellcolor[HTML]{DBDBDB}\textbf{4} & 2.29195 & 1.76471 & 1.33492 & -0.40441636, -0.01719112 & 0.00343645, 0.49993767 & 79 & 61 & 0.98787 & 0.98714 \\ \hline
\cellcolor[HTML]{DBDBDB}\textbf{5} & 2.42279 & 3.18159 & 5.57723 & -0.50313864, -0.03521012 & 0.04263622, 0.33996168 & 85 & 68 & 0.98565 & 0.98659 \\ \hline
\cellcolor[HTML]{DBDBDB}\textbf{6} & 2.07221 & 2.36239 & 5.79011 & -0.419110756, -0.261993043 & 2.76648[-06], 0.256172562 & 71 & 59 & 0.98885 & 0.98613 \\ \hline
\cellcolor[HTML]{DBDBDB}\textbf{7} & 2.06019 & 1.67036 & 2.73084 & -0.302222205, -0.278906844 & -1.01643954[-20], 0.234360342 & 65 & 57 & 0.98565 & 0.98714 \\ \hline
\cellcolor[HTML]{DBDBDB}\textbf{8} & 2.96621 & 2.51154 & 2.51168 & -0.473367622, -0.413235912 & -5.55111512[-21], 0.487328526 & 88 & 35 & 0.98883 & 0.99666 \\ \hline
\cellcolor[HTML]{DBDBDB}\textbf{9} & 2.26002 & 1.81584 & 4.45387 & -0.55791953, -0.0460368707 & 2.85695[-05], 0.475565246 & 71 & 61 & 0.98673 & 0.98455 \\ \hline
\cellcolor[HTML]{DBDBDB}\textbf{10} & 1.35691 & 0.61534 & 5.02339 & -0.282913602, -0.0557603837 & 0.000208365485, 0.5 & 87 & 59 & 0.98860 & 0.98595 \\ \hline
\cellcolor[HTML]{DBDBDB}\textbf{11} & 2.52209 & 1.73075 & 0.40260 & -0.23743752, -0.16350079 & 0.00959416, 0.32915802 & 83 & 61 & 0.98896 & 0.98652 \\ \hline
\cellcolor[HTML]{DBDBDB}\textbf{12} & 2.74583 & 2.56126 & 3.80340 & -0.51353766, -0.0883247645 & 1.03362794[-21], 0.411686847 & 82 & 64 & 0.98887 & 0.98725 \\ \hline
\cellcolor[HTML]{DBDBDB}\textbf{13} & 3.10688 & 2.83284 & 5.32182 & -0.46327894, -0.151392808 & 8.9073[-05], 0.333453602 & 90 & 65 & 0.98894 & 0.98815 \\ \hline
\cellcolor[HTML]{DBDBDB}\textbf{14} & 1.95288 & 4.53352 & 3.46380 & -0.433075568, -0.135160185 & 0.000326574362, 0.473647838 & 71 & 56 & 0.98534 & 0.98521 \\ \hline
\cellcolor[HTML]{DBDBDB}\textbf{15} & 2.87028 & 1.87002 & 3.94885 & -0.21353527, -0.0947253 & 0.11645518, 0.49489148 & 88 & 66 & 0.98923 & 0.98786 \\ \hline
\cellcolor[HTML]{DBDBDB}\textbf{16} & 2.15739 & 3.07773 & 5.71602 & -0.25089681, -0.08538334 & 0.00192315, 0.46304837 & 78 & 61 & 0.98881 & 0.98542 \\ \hline
\cellcolor[HTML]{DBDBDB}\textbf{17} & 2.79398 & 5.42806 & 6.24226 & -0.307971979, -0.00695713502 & 0.000188672339, 0.499968525 & 86 & 60 & 0.98941 & 0.98679 \\ \hline
\cellcolor[HTML]{DBDBDB}\textbf{18} & 1.78857 & 0.76770 & 2.56039 & -0.520485426, -0.146095907 & 2.08166817[-20], 0.374574699 & 71 & 61 & 0.98826 & 0.98721 \\ \hline
\cellcolor[HTML]{DBDBDB}\textbf{19} & 2.68895 & 0.81678 & 2.57334 & -0.468694228, -1[-10] & 8.30025[-05], 0.5 & 89 & 63 & 0.98873 & 0.98800 \\ \hline
\cellcolor[HTML]{DBDBDB}\textbf{20} & 2.50068 & 1.42982 & 5.42760 & -0.564136098, -0.175077687 & -2.70483796[-21], 0.374246631 & 82 & 66 & 0.98733 & 0.98744 \\ \hline
\cellcolor[HTML]{DBDBDB}\textbf{21} & 2.38126 & 1.40436 & 5.64973 & -0.406232817, -9.85128[-05] & -5.55111512[-21], 0.5 & 79 & 59 & 0.98672 & 0.98627 \\ \hline
\cellcolor[HTML]{DBDBDB}\textbf{22} & 2.14670 & 1.98890 & 5.87210 & -0.364938465, -0.120608727 & 0.000128319093, 0.435746907 & 75 & 58 & 0.98755 & 0.98407 \\ \hline
\cellcolor[HTML]{DBDBDB}\textbf{23} & 2.76371 & 5.67223 & 5.18676 & -0.41536678, -0.0460577 & 0.00844998, 0.49991171 & 87 & 62 & 0.98898 & 0.98857 \\ \hline
\cellcolor[HTML]{DBDBDB}\textbf{24} & 2.35484 & 2.12405 & 5.77071 & -0.2572922, -0.224169 & 0.01555949, 0.43364061 & 83 & 61 & 0.98793 & 0.98660 \\ \hline
\cellcolor[HTML]{DBDBDB}\textbf{25} & 2.14131 & 3.88017 & 4.43203 & -0.929939457, -0.369182991 & -9.7584383[-17], 0.0000850766777 & 76 & 60 & 0.98763 & 0.98572 \\ \hline
\cellcolor[HTML]{DBDBDB}\textbf{26} & 2.70824 & 1.78139 & 4.29155 & -0.32973799, -0.13942659 & 0.13863571, 0.41938758 & 84 & 62 & 0.98851 & 0.98509 \\ \hline
\cellcolor[HTML]{DBDBDB}\textbf{27} & 1.97566 & 2.68021 & 0.50074 & -0.383118824, -2[-10] & -8.14579432[-20], 0.499902518 & 76 & 61 & 0.98894 & 0.98525 \\ \hline
\cellcolor[HTML]{DBDBDB}\textbf{28} & 1.73206 & 0.07675 & 2.44859 & -0.388221058, -0.0573922577 & -5.93648842[-21], 0.477696285 & 70 & 54 & 0.98636 & 0.98677 \\ \hline
\cellcolor[HTML]{DBDBDB}\textbf{29} & 3.11359 & 2.04632 & 4.96774 & -0.29776454, -0.22670669 & 0.00052851, 0.31971397 & 90 & 59 & 0.98862 & 0.98815 \\ \hline
\cellcolor[HTML]{DBDBDB}\textbf{30} & 2.20927 & 3.00257 & 3.56097 & -0.424833383, -0.284910373 & -2.74925531[-21], 0.18220541 & 74 & 57 & 0.98420 & 0.98325 \\ \hline
\cellcolor[HTML]{DBDBDB}\textbf{31} & 2.12776 & 6.08943 & 4.49468 & -0.845467863, -1[-10] & 6.77626358[-20], 0.496124243 & 78 & 68 & 0.98737 & 0.98766 \\ \hline
\cellcolor[HTML]{DBDBDB}\textbf{32} & 2.85530 & 4.67939 & 4.28628 & -0.337961514, -0.0986680445 & 0.000138723458, 0.5 & 86 & 61 & 0.98848 & 0.98806 \\ \hline
\cellcolor[HTML]{DBDBDB}\textbf{33} & 2.41524 & 3.79065 & 3.00641 & -0.432054855, -0.403254422 & 4.536599[-05], 0.242284549 & 77 & 64 & 0.98751 & 0.98602 \\ \hline
\cellcolor[HTML]{DBDBDB}\textbf{34} & 2.93550 & 5.17916 & 4.46644 & -0.34218794, -0.1422061 & 0.0039324, 0.24517162 & 84 & 62 & 0.98941 & 0.98724 \\ \hline
\cellcolor[HTML]{DBDBDB}\textbf{35} & 2.69760 & 4.06237 & 2.82939 & -0.618683055, -0.390404886 & -6.19274316[-21], 6936589[-05] & 87 & 61 & 0.98868 & 0.98717 \\ \hline
\cellcolor[HTML]{DBDBDB}\textbf{36} & 2.26660 & 4.60415 & 4.39647 & -0.45970215, -0.13173758 & 0.00082242, 0.42017257 & 80 & 63 & 0.98841 & 0.98717 \\ \hline
\cellcolor[HTML]{DBDBDB}\textbf{37} & 2.23500 & 1.56288 & 2.13387 & -0.45002646, -0.00158542 & 0.00091409, 0.49977423 & 79 & 60 & 0.98897 & 0.98729 \\ \hline
\cellcolor[HTML]{DBDBDB}\textbf{38} & 2.77113 & 4.38132 & 1.28333 & -0.613252871, -0.317785931 & 6.20192605[-21], 0.0110331927 & 81 & 61 & 0.98847 & 0.98718 \\ \hline
\cellcolor[HTML]{DBDBDB}\textbf{39} & 2.88243 & 5.35496 & 0.00254 & -0.428712027, -9.99999034[-11] & 2.42861287[-21], 0.5 & 85 & 61 & 0.98904 & 0.98841 \\ \hline
\cellcolor[HTML]{DBDBDB}\textbf{40} & 2.37142 & 3.14858 & 1.06018 & -0.499436305, -0.000161631827 & 3.432516[-05], 0.393702916 & 80 & 65 & 0.98779 & 0.98525 \\ \hline
\cellcolor[HTML]{DBDBDB}\textbf{41} & 2.21118 & 4.30358 & 5.56859 & -0.39081709, -0.24157334 & 0.0050181, 0.19390617 & 77 & 59 & 0.98820 & 0.98798 \\ \hline
\cellcolor[HTML]{DBDBDB}\textbf{42} & 2.76569 & 4.78934 & 3.68815 & -0.27655431, -0.09309335 & 0.05960735, 0.41938602 & 88 & 66 & 0.98993 & 0.98751 \\ \hline
\cellcolor[HTML]{DBDBDB}\textbf{43} & 2.44340 & 5.86060 & 0.66880 & -0.78475823, -0.49585084 & 0, 0.05258665 & 81 & 65 & 0.98731 & 0.98699 \\ \hline
\cellcolor[HTML]{DBDBDB}\textbf{44} & 2.34349 & 5.17034 & 3.42800 & -0.99856133, -0.50502313 & 0.00258655, 0.00518738 & 80 & 62 & 0.98899 & 0.98745 \\ \hline
\cellcolor[HTML]{DBDBDB}\textbf{45} & 2.89725 & 3.25978 & 2.46404 & -0.387278318, -3.44764[-06] & -5.58731656[-19], 0.5 & 88 & 59 & 0.98871 & 0.98685 \\ \hline
\cellcolor[HTML]{DBDBDB}\textbf{46} & 2.22109 & 3.17582 & 4.57342 & -0.500541647, -0.244572891 & -7.09658898[-21], 0.097607618 & 76 & 60 & 0.98709 & 0.98540 \\ \hline
\cellcolor[HTML]{DBDBDB}\textbf{47} & 2.12925 & 5.74234 & 4.16431 & -0.771209557, -0.511318543 & 6.77593[-05], 0.011359187 & 78 & 61 & 0.98965 & 0.98821 \\ \hline
\cellcolor[HTML]{DBDBDB}\textbf{48} & 2.04236 & 4.02405 & 4.75088 & -0.36012526, -0.08207992 & 0.00086559, 0.5 & 77 & 59 & 0.98810 & 0.98672 \\ \hline
\cellcolor[HTML]{DBDBDB}\textbf{49} & 2.84364 & 3.51744 & 2.31714 & -0.491466218, -0.210036697 & 1.15012[-05] 0.15350186 & 87 & 62 & 0.98917 & 0.98769 \\ \hline
\cellcolor[HTML]{DBDBDB}\textbf{50} & 2.82916 & 1.62803 & 2.14790 & -0.64864277, -0.24673651 & 0.00116572, 0.2122708 & 87 & 66 & 0.98852 & 0.98784 \\ \hline
\cellcolor[HTML]{DBDBDB}\textbf{51} & 2.89071 & 3.02084 & 4.21549 & -0.546182131, -0.253262929 & 8.822617[-05], 0.132792954 & 88 & 61 & 0.98970 & 0.98744 \\ \hline
\cellcolor[HTML]{DBDBDB}\textbf{52} & 2.46587 & 2.72146 & 4.64330 & -0.297280218, -2[-10] & 9.43813[-05], 0.5 & 84 & 61 & 0.98733 & 0.98507 \\ \hline
\cellcolor[HTML]{DBDBDB}\textbf{53} & 2.08611 & 1.21178 & 0.73849 & -0.4268411, -0.19888518 & 0, 0.5 & 76 & 59 & 0.98655 & 0.98704 \\ \hline
\cellcolor[HTML]{DBDBDB}\textbf{54} & 2.32228 & 2.97413 & 0.49990 & -0.40492888, -0.02533541 & 0.03269568, 0.49191016 & 80 & 63 & 0.98892 & 0.98833 \\ \hline
\cellcolor[HTML]{DBDBDB}\textbf{55} & 2.58994 & 5.10622 & 0.32334 & -0.39498385, -0.00474822 & 0.00495389, 0.49998633 & 85 & 61 & 0.98805 & 0.98648 \\ \hline
\cellcolor[HTML]{DBDBDB}\textbf{56} & 3.12728 & 2.66350 & 0.39019 & -0.326699117, -0.0739915233 & -9.20043811[-22], 0.339481137 & 87 & 65 & 0.98958 & 0.98674 \\ \hline
\cellcolor[HTML]{DBDBDB}\textbf{57} & 2.47000 & 5.43223 & 0.54154 & -0.482205587, -7.549212[-05] & -5.0491846[-18], 0.5 & 80 & 60 & 0.98714 & 0.98682 \\ \hline
\cellcolor[HTML]{DBDBDB}\textbf{58} & 2.72399 & 3.71018 & 6.01040 & -0.303331314, -0.150786279 & 1.92493278[-20], 0.499955936 & 84 & 60 & 0.98957 & 0.98678 \\ \hline
\cellcolor[HTML]{DBDBDB}\textbf{59} & 2.90324 & 4.67714 & 1.73488 & -0.625136304, -0.281968253 & 5.55111512[-21], 0.00316742482 & 90 & 65 & 0.98831 & 0.98649 \\ \hline
\cellcolor[HTML]{DBDBDB}\textbf{60} & 2.99424 & 4.13582 & 4.91289 & -0.560501634, -0.119668925 & 3.575767[-05], 0.360405124 & 88 & 65 & 0.98958 & 0.98844 \\ \hline
\cellcolor[HTML]{DBDBDB}\textbf{61} & 2.04922 & 5.46014 & 5.29590 & -0.639314505, -0.411878255 & 2.87362[-05], 0.0798711858 & 74 & 61 & 0.98676 & 0.98593 \\ \hline
\cellcolor[HTML]{DBDBDB}\textbf{62} & 2.41078 & 2.27029 & 2.80876 & -0.385793183, -0.0907787454 & 2.74816[-05], 0.499943768 & 81 & 60 & 0.98786 & 0.98726 \\ \hline
\cellcolor[HTML]{DBDBDB}\textbf{63} & 2.21419 & 3.89046 & 5.83689 & -0.317419452, -0.224085882 & 3.83378[-06], 0.0831143443 & 75 & 63 & 0.98874 & 0.98588 \\ \hline
\cellcolor[HTML]{DBDBDB}\textbf{64} & 2.94148 & 1.41033 & 4.49399 & -0.270611167, -0.212768941 & 1.12237808[-20], 0.478865876 & 84 & 52 & 0.98873 & 0.98986 \\ \hline
\cellcolor[HTML]{DBDBDB}\textbf{65} & 2.15569 & 3.79110 & 0.80333 & -0.658249637, -0.0313312003 & 1.27054942[-21], 0.367934754 & 76 & 62 & 0.98641 & 0.98578 \\ \hline
\cellcolor[HTML]{DBDBDB}\textbf{66} & 2.12855 & 6.09186 & 5.37105 & -0.344879199, -0.273737253 & 3.83749[-05], 0.431750646 & 75 & 56 & 0.98511 & 0.98549 \\ \hline
\cellcolor[HTML]{DBDBDB}\textbf{67} & 2.46317 & 5.92169 & 0.92723 & -0.9999999999, -0.350759059 & -5.20901191[-20], 0.000227563952 & 82 & 60 & 0.98751 & 0.98583 \\ \hline
\cellcolor[HTML]{DBDBDB}\textbf{68} & 2.57193 & 3.08012 & 2.47826 & -0.30641164, -0.23576931 & 0.00577656, 0.29877037 & 82 & 58 & 0.98741 & 0.98741 \\ \hline
\cellcolor[HTML]{DBDBDB}\textbf{69} & 2.96948 & 0.51671 & 4.13995 & -0.969678914, -0.318876434 & 5.55111512[-21], 0.00696480632 & 87 & 62 & 0.98854 & 0.98628 \\ \hline
\cellcolor[HTML]{DBDBDB}\textbf{70} & 2.70184 & 0.12106 & 2.01037 & -0.409739807, -0.0117086249 & 2.22044605[-20], 0.498682579 & 83 & 59 & 0.98916 & 0.98827 \\ \hline
\cellcolor[HTML]{DBDBDB}\textbf{71} & 2.66267 & 0.41965 & 3.55148 & -0.33466489, -0.01850706 & 0.04845566, 0.49860204 & 84 & 63 & 0.98930 & 0.98762 \\ \hline
\cellcolor[HTML]{DBDBDB}\textbf{72} & 2.98978 & 1.38898 & 1.96907 & -0.40022605, -1[-10] & -6.16297582[-33], 0.499910998 & 86 & 66 & 0.98937 & 0.98676 \\ \hline
\cellcolor[HTML]{DBDBDB}\textbf{73} & 2.75605 & 3.05189 & 4.90998 & -0.75059603, -1[-10] & 2.02187991[-21], 0.442531917 & 87 & 66 & 0.98894 & 0.98796 \\ \hline
\cellcolor[HTML]{DBDBDB}\textbf{74} & 2.91441 & 4.72570 & 0.86171 & -0.45636805, -0.07078021 & 0.0083545, 0.43573367 & 86 & 63 & 0.98928 & 0.98744 \\ \hline
\cellcolor[HTML]{DBDBDB}\textbf{75} & 2.27941 & 5.09299 & 3.50890 & -0.31793731, -0.13291007 & 0.00156619, 0.46931166 & 76 & 57 & 0.98831 & 0.98695 \\ \hline
\cellcolor[HTML]{DBDBDB}\textbf{76} & 2.72950 & 3.56458 & 0.87768 & -0.234274471, -0.221418761 & 6.227788[-05], 0.499907965 & 87 & 53 & 0.98819 & 0.98787 \\ \hline
\cellcolor[HTML]{DBDBDB}\textbf{77} & 2.76295 & 5.11002 & 0.06402 & -0.282738656, -0.1626622 & 5.0062[-05], 0.32528278 & 85 & 61 & 0.98875 & 0.98703 \\ \hline
\cellcolor[HTML]{DBDBDB}\textbf{78} & 2.72094 & 1.39060 & 3.53269 & -0.423759495, -0.310603864 & 9.50142[-05], 0.113009692 & 87 & 62 & 0.98924 & 0.98794 \\ \hline
\cellcolor[HTML]{DBDBDB}\textbf{79} & 2.99682 & 3.00681 & 2.64991 & -0.231924605, -0.177429066 & -2.63787115[-20], 0.5 & 87 & 53 & 0.98861 & 0.98860 \\ \hline
\cellcolor[HTML]{DBDBDB}\textbf{80} & 2.86026 & 4.39329 & 3.54436 & -0.582593075, -0.362445833 & 4.50945[-06], 0.00342749573 & 83 & 61 & 0.98938 & 0.98795 \\ \hline
\cellcolor[HTML]{DBDBDB}\textbf{81} & 2.39940 & 0.21368 & 1.29938 & -0.32610197, -0.268081307 & 1.29557[-05], 0.5 & 82 & 64 & 0.98822 & 0.98711 \\ \hline
\cellcolor[HTML]{DBDBDB}\textbf{82} & 1.89477 & 2.81396 & 0.23890 & -0.34113826, -0.04654715 & 0.02214931, 0.49999074 & 75 & 57 & 0.98761 & 0.98719 \\ \hline
\cellcolor[HTML]{DBDBDB}\textbf{83} & 2.18455 & 3.97272 & 1.39291 & -0.283828395, -0.201691582 & 0.00021136743, 0.362781599 & 80 & 59 & 0.98474 & 0.98371 \\ \hline
\cellcolor[HTML]{DBDBDB}\textbf{84} & 2.49180 & 2.58583 & 5.14075 & -0.653333749, -0.383343467 & 1.56079[-05], 0.0658074559 & 81 & 63 & 0.98895 & 0.98740 \\ \hline
\cellcolor[HTML]{DBDBDB}\textbf{85} & 2.59845 & 2.37573 & 3.12417 & -0.621529941, -0.56172392 & 0.000137961341, 0.00064905387 & 84 & 66 & 0.98824 & 0.98825 \\ \hline
\cellcolor[HTML]{DBDBDB}\textbf{86} & 2.60883 & 6.17957 & 2.03449 & -0.33353985, -0.00353956 & 0.00062256, 0.49998387 & 85 & 59 & 0.98831 & 0.98776 \\ \hline
\cellcolor[HTML]{DBDBDB}\textbf{87} & 2.85977 & 3.13459 & 4.87675 & -0.735734897, -0.521398633 & -2.84750996[-20], 0.0000855691489 & 83 & 63 & 0.98925 & 0.98889 \\ \hline
\cellcolor[HTML]{DBDBDB}\textbf{88} & 3.05267 & 2.09009 & 0.15869 & -0.439726034, -0.276376103 & 8.36074[-05], 0.323169545 & 88 & 66 & 0.98953 & 0.98892 \\ \hline
\cellcolor[HTML]{DBDBDB}\textbf{89} & 2.65536 & 1.53336 & 2.01915 & -0.5956152, -0.1775187 & 0.00273495, 0.26232643 & 83 & 64 & 0.98950 & 0.98805 \\ \hline
\cellcolor[HTML]{DBDBDB}\textbf{90} & 2.42194 & 5.35202 & 1.63601 & -0.713680279, -0.32358958 & -2.11556797[-20], 0.0460521672 & 82 & 61 & 0.98628 & 0.98475 \\ \hline
\cellcolor[HTML]{DBDBDB}\textbf{91} & 2.16767 & 1.66413 & 1.52262 & -0.38479997, -0.174828392 & 1.11022302[-20], 0.428639597 & 80 & 60 & 0.98895 & 0.98738 \\ \hline
\cellcolor[HTML]{DBDBDB}\textbf{92} & 2.98966 & 3.05187 & 1.47806 & -0.42503145, -0.1299089 & 0.0034686, 0.33280943 & 86 & 63 & 0.98810 & 0.98760 \\ \hline
\cellcolor[HTML]{DBDBDB}\textbf{93} & 2.80563 & 4.43216 & 0.93327 & -0.467989285, -9.99999736[-11] & 1.77245422[-21], 0.499900008 & 88 & 64 & 0.98872 & 0.98726 \\ \hline
\cellcolor[HTML]{DBDBDB}\textbf{94} & 2.76116 & 3.67330 & 3.58557 & -0.483856094, -6.68692[-05] & -5.18797705[-18], 0.49997342 & 88 & 62 & 0.98863 & 0.98729 \\ \hline
\cellcolor[HTML]{DBDBDB}\textbf{95} & 2.86817 & 4.82626 & 0.79992 & -0.45313733, -2[-10] & 3.69508592[-19], 0.499900763 & 86 & 65 & 0.98895 & 0.98641 \\ \hline
\cellcolor[HTML]{DBDBDB}\textbf{96} & 3.10749 & 6.05200 & 0.55580 & -0.33922282, -0.0857602356 & 8.77796[-05], 0.456862679 & 89 & 59 & 0.98866 & 0.98790 \\ \hline
\cellcolor[HTML]{DBDBDB}\textbf{97} & 2.52530 & 1.56729 & 5.60700 & -0.32663141, -0.02261459 & 0.00403472, 0.5 & 81 & 63 & 0.98761 & 0.98575 \\ \hline
\cellcolor[HTML]{DBDBDB}\textbf{98} & 2.61797 & 5.00611 & 4.61394 & -0.32243556, -0.12939632 & 0.00912948, 0.43418299 & 84 & 62 & 0.98958 & 0.98806 \\ \hline
\cellcolor[HTML]{DBDBDB}\textbf{99} & 2.53796 & 3.96390 & 1.80047 & -0.26281421, -0.16513112 & 0.0335583, 0.49467932 & 78 & 61 & 0.98766 & 0.98604 \\ \hline
\cellcolor[HTML]{DBDBDB}\textbf{100} & 2.31079 & 5.94527 & 1.46166 & -0.50568839, -0.1482959 & 0.00892103, 0.45501687 & 78 & 63 & 0.98772 & 0.98645 \\ \hline
\caption{The data for learning speed optimization.}
\label{table1}
\end{longtable*}

We can note that in this case the fidelity obtained with the optimal PDF is almost the same that without optimization because we are not optimizing de accuracy of the result, but the number of iterations is reduce in more than $20\%$ in average. 

\newpage
\subsection{Learning speed Histogram}
\begin{figure}[t]
\centering
\includegraphics[width=1\linewidth]{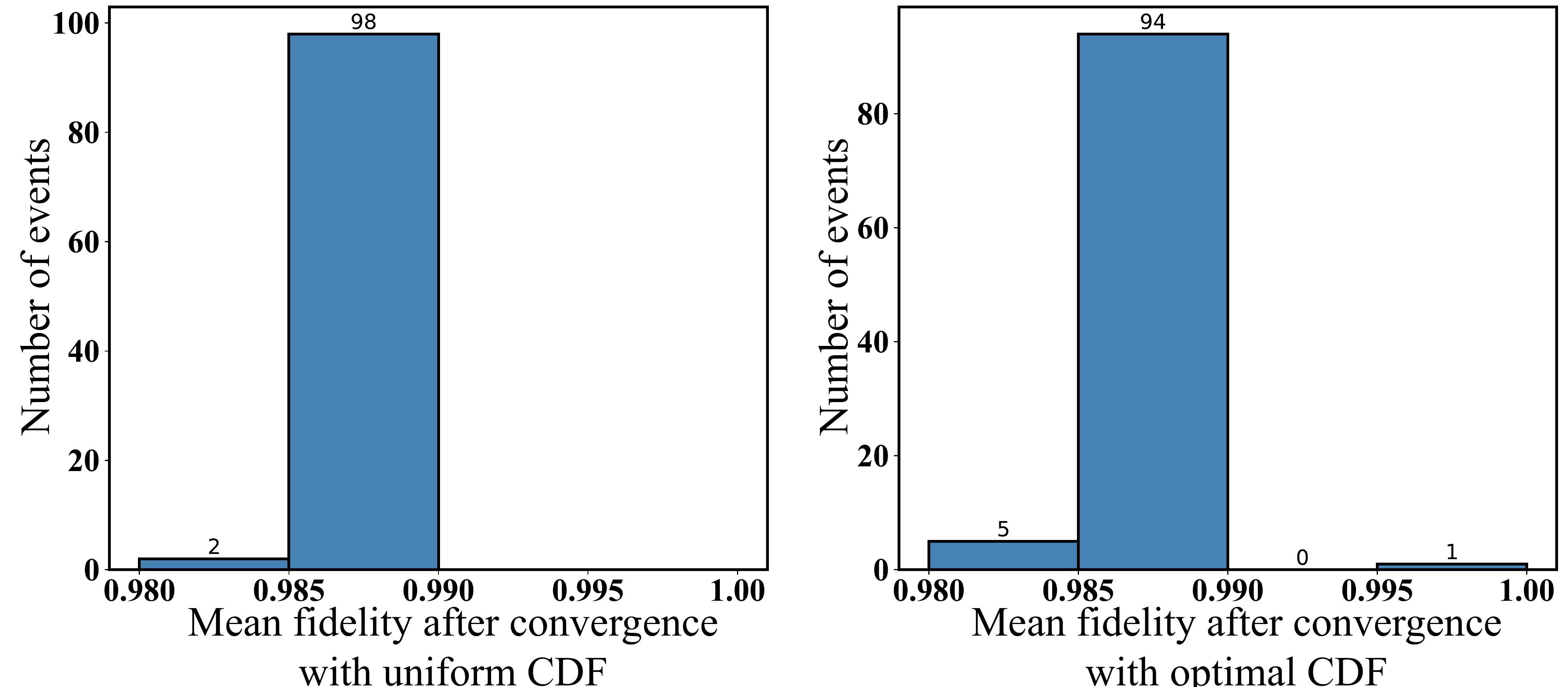}
\caption{Histogram for mean fidelity without (left panel) and with optimization (right panel) for 100 different $U(\theta,\phi,\lambda)$, for the optimization of learning speed.}
\label{Fig04}
\end{figure}

In the figure~\ref{Fig04}, we summarize the data obtained in table~\ref{table1}. Left panel shows the data without optimal PDF, and right panel shows the data with optimal PDF. We note that both histogram are basically the same, which means that the learning speed optimization will not affect the fidelity or accuracy of the final result.

\newpage
\section{Learning accuracy data}

The next table collect all the data for the 100 different instances for the choose of $\{\theta,\phi,\lambda\}$ and their  $X_in$ and $Y_in$ for the optimal PCF. Also the table shows the fidelity for the optimal ($F_{opt}$) and non-optimal PDF ($F$) as well as the number of iteration used ($N$). In this case we are optimizing the learning accurancy.

\begin{longtable*}{
>{\columncolor[HTML]{D4D4D4}}l |l|l|l|l|l|l|l|l|}
\hline
\cellcolor[HTML]{B0B3B2} & \cellcolor[HTML]{B0B3B2}\bm{$\theta$} & \cellcolor[HTML]{B0B3B2}\bm{$\phi$} & \cellcolor[HTML]{B0B3B2}\bm{$\lambda$} & \cellcolor[HTML]{B0B3B2}$X_{in}$ & \cellcolor[HTML]{B0B3B2}$Y_{in}$ & \cellcolor[HTML]{B0B3B2}$N$ & \cellcolor[HTML]{B0B3B2}$F$ & \cellcolor[HTML]{B0B3B2}$F_{opt}$ \\ \hline
\textbf{1} & 2.83651 & 2.51483 & 5.79311 & -0.22658129, -0.02628605 & 6.46642309[-22], 0.499681951 & 80 & 0.95489 & 0.96506 \\ \hline
\textbf{2} & 2.60239 & 2.91385 & 1.94757 & -0.38593267, -0.06560765 & 4.99435[-05], 0.498704474 & 80 & 0.95075 & 0.97406 \\ \hline
\textbf{3} & 1.62294 & 2.66070 & 0.16163 & -0.602975987, -1[-10] & 0.01540395, 0.49991607 & 80 & 0.97004 & 0.97565 \\ \hline
\textbf{4} & 2.29195 & 1.76471 & 1.33492 & -0.52017218, -0.33360453 & 3.87541[-05], 0.211416718 & 80 & 0.95621 & 0.97061 \\ \hline
\textbf{5} & 2.42279 & 3.18159 & 5.57723 & -0.35962401, -0.07388651 & 0.02790605, 0.5 & 80 & 0.95731 & 0.97164 \\ \hline
\textbf{6} & 2.07221 & 2.36239 & 5.79011 & -0.78203655, -0.29764012 & -1.28986258[-20], 0.0923301799 & 80 & 0.95675 & 0.96929 \\ \hline
\textbf{7} & 2.06019 & 1.67036 & 2.73084 & -0.37643844, -0.01952776 & 4.5851[-06], 0.499960074 & 80 & 0.96556 & 0.97080 \\ \hline
\textbf{8} & 2.96621 & 2.51154 & 2.51168 & -0.98231875, -0.50135792 & 0.00018449, 0.00037358 & 80 & 0.94750 & 0.97031 \\ \hline
\textbf{9} & 2.26002 & 1.81584 & 4.45387 & -0.98949422, -0.5671099 & 0.01415337, 0.03809302 & 80 & 0.95098 & 0.96492 \\ \hline
\textbf{10} & 1.35691 & 0.61534 & 5.02339 & -0.29599157, -0.57327287 & 0.31984894, 0.4923946 & 80 & 0.93653 & 0.94413 \\ \hline
\textbf{11} & 2.87028 & 1.87002 & 3.94885 & -0.45664744, -0.29318509 & 0.0095358, 0.05159221 & 80 & 0.94830 & 0.96723 \\ \hline
\textbf{12} & 2.15739 & 3.07773 & 5.71602 & -0.42998516, -0.24883537 & 0.13140672, 0.29338095 & 80 & 0.95490 & 0.97120 \\ \hline
\textbf{13} & 2.52209 & 1.73075 & 0.40260 & -0.31623152, -0.14464535 & 0.01684554, 0.33745235 & 80 & 0.95596 & 0.97104 \\ \hline
\textbf{14} & 2.74583 & 2.56126 & 3.80340 & -0.50005984, -0.12676685 & 0.00346618, 0.21540484 & 80 & 0.95053 & 0.96758 \\ \hline
\textbf{15} & 3.10688 & 2.83284 & 5.32182 & -0.264065269, -9.999[-11] & 0, 0.5 & 80 & 0.95043 & 0.96972 \\ \hline
\textbf{16} & 1.95288 & 4.53352 & 3.46380 & -0.388548,    -0.17632076 & 0.01503028, 0.38622965 & 80 & 0.96176 & 0.97168 \\ \hline
\textbf{17} & 1.73206 & 0.07675 & 2.44859 & -0.28327408, -0.26619532 & 0.09564394, 0.35047785 & 80 & 0.96741 & 0.97520 \\ \hline
\textbf{18} & 2.20927 & 3.00257 & 3.56097 & -0.32950066, -0.26883062 & 0.02256158, 0.28228575 & 80 & 0.95314 & 0.96663 \\ \hline
\textbf{19} & 2.79398 & 5.42806 & 6.24226 & -0.95830604, -0.38431026 & 0.00661281, 0.00880922 & 80 & 0.95065 & 0.97053 \\ \hline
\textbf{20} & 2.38126 & 1.40436 & 5.64973 & -0.9554078,   -0.43829696 & 0.00012443, 0.02216138 & 80 & 0.95351 & 0.96658 \\ \hline
\textbf{21} & 2.50068 & 1.42982 & 5.42760 & -0.50448542, -0.10994467 & -6.77626358[-21], 0.49995152 & 80 & 0.95049 & 0.96533 \\ \hline
\textbf{22} & 1.78857 & 0.76770 & 2.56039 & -0.0624920122, -3.6113[-05] & 0.37509291, 0.5 & 80 & 0.96327 & 0.97092 \\ \hline
\textbf{23} & 2.68895 & 0.81678 & 2.57334 & -0.251802237, -1[-10] & 0.07207826, 0.5 & 80 & 0.95488 & 0.96392 \\ \hline
\textbf{24} & 2.76371 & 5.67223 & 5.18676 & -0.20446587, -0.19910618 & -5.55111512[-21], 0.499977945 & 80 & 0.95454 & 0.97839 \\ \hline
\textbf{25} & 2.35484 & 2.12405 & 5.77071 & -0.79739122, -0.43221379 & 0.00517663, 0.06708963 & 80 & 0.95573 & 0.96874 \\ \hline
\textbf{26} & 2.14670 & 1.98890 & 5.87210 & -0.22988712, -0.16913267 & 0.000041023229, 0.5 & 80 & 0.95506 & 0.97596 \\ \hline
\textbf{27} & 2.14131 & 3.88017 & 4.43203 & -0.99994708, -0.50247325 & -5.36055763[-20], 0.00700035073 & 80 & 0.96013 & 0.97200 \\ \hline
\textbf{28} & 2.70824 & 1.78139 & 4.29155 & -0.26922985, -0.07666227 & 0.0000662211987, 0.5 & 80 & 0.94925 & 0.97055 \\ \hline
\textbf{29} & 3.11359 & 2.04632 & 4.96774 & -0.49732798, -0.01685642 & 0.0000174612887, 0.287787822 & 80 & 0.95234 & 0.96586 \\ \hline
\textbf{30} & 1.97566 & 2.68021 & 0.50074 & -0.698348786, -8.94312[-05] & 1.96800135[-20], 0.347860102 & 80 & 0.96010 & 0.97122 \\ \hline
\textbf{31} & 2.12776 & 6.08943 & 4.49468 & -0.6823936,   -0.43406705 & -1.49253285[-20], 8.2134[-05] & 80 & 0.96315 & 0.97325 \\ \hline
\textbf{32} & 2.85530 & 4.67939 & 4.28628 & -0.91867228, -0.30253779 & 1.552483[-05],6.98294[-05] & 80 & 0.95623 & 0.97096 \\ \hline
\textbf{33} & 2.41524 & 3.79065 & 3.00641 & -0.37244503, -0.34949234 & 0.00355155, 0.04073554 & 80 & 0.95088 & 0.96893 \\ \hline
\textbf{34} & 2.93550 & 5.17916 & 4.46644 & -0.65012892, -0.00071334 & 9.920904[-05], 0.499348973 & 80 & 0.95534 & 0.96503 \\ \hline
\textbf{35} & 2.69760 & 4.06237 & 2.82939 & -0.63380379, -0.49980422 & 2.02733247[-21], 0.0145546481 & 80 & 0.95429 & 0.96918 \\ \hline
\textbf{36} & 2.26660 & 4.60415 & 4.39647 & -0.113053973, -2[-10] & 0.36858133, 0.4999619 & 80 & 0.95881 & 0.96739 \\ \hline
\textbf{37} & 2.23500 & 1.56288 & 2.13387 & -0.45531637, -0.00497668 & 0.00062212, 0.49991192 & 80 & 0.95726 & 0.97510 \\ \hline
\textbf{38} & 2.77113 & 4.38132 & 1.28333 & -0.24967158, -0.15935152 & -1.11022302[-20], 0.328961357 & 80 & 0.95507 & 0.97155 \\ \hline
\textbf{39} & 2.88243 & 5.35496 & 0.00254 & -0.427704684, -2[-10] & 0, 0.49990047 & 80 & 0.94829 & 0.96905 \\ \hline
\textbf{40} & 2.37142 & 3.14858 & 1.06018 & -0.4651497, -0.126999 & 0.000245935043, 0.454376543 & 80 & 0.95579 & 0.97436 \\ \hline
\textbf{41} & 2.21118 & 4.30358 & 5.56859 & -0.46724996, -0.14351635 & 9.56351687[-21], 0.387786874 & 80 & 0.96100 & 0.97322 \\ \hline
\textbf{42} & 2.76569 & 4.78934 & 3.68815 & -0.716947108, -9.9999[-11] & 0.0000781179353, 0.38177391 & 80 & 0.94966 & 0.96441 \\ \hline
\textbf{43} & 2.44340 & 5.86060 & 0.66880 & -0.04856462, -0.01146986 & 0.39178085, 0.5 & 80 & 0.95449 & 0.96311 \\ \hline
\textbf{44} & 2.34349 & 5.17034 & 3.42800 & -0.50962221, -0.10198462 & -2.44669142[-21], 0.321459154 & 80 & 0.95993 & 0.97187 \\ \hline
\textbf{45} & 2.89725 & 3.25978 & 2.46404 & -0.24073117, -0.14868635 & 0.23972925, 0.48865284 & 80 & 0.95361 & 0.96456 \\ \hline
\textbf{46} & 2.22109 & 3.17582 & 4.57342 & -0.62021665, -0.40811444 & -4.32608973[-22], 0.13008986 & 80 & 0.96011 & 0.96776 \\ \hline
\textbf{47} & 2.12925 & 5.74234 & 4.16431 & -0.9871146,   -0.67280069 & 2.0287[-05], 7.73143[-05] & 80 & 0.95991 & 0.96988 \\ \hline
\textbf{48} & 2.04236 & 4.02405 & 4.75088 & -0.68862154, -0.01168985 & 6.60527[-05], 0.499981325 & 80 & 0.96238 & 0.96944 \\ \hline
\textbf{49} & 2.84364 & 3.51744 & 2.31714 & -0.32325424, -0.14378023 & 3.83964[-05], 0.499543956 & 80 & 0.94615 & 0.96817 \\ \hline
\textbf{50} & 2.82916 & 1.62803 & 2.14790 & -0.37096935, -0.00577171 & -1.47512731[-21], 0.5 & 80 & 0.95368 & 0.97311 \\ \hline
\textbf{51} & 2.89071 & 3.02084 & 4.21549 & -0.29921219, -0.00906005 & -1.35525272[-20], 0.499257551 & 80 & 0.95582 & 0.97193 \\ \hline
\textbf{52} & 2.46587 & 2.72146 & 4.64330 & -0.36316915, -0.17747838 & 0.04782131, 0.32694109 & 80 & 0.95432 & 0.96942 \\ \hline
\textbf{53} & 2.08611 & 1.21178 & 0.73849 & -0.499964682, -9.99999451[-11] & -1.79986232[-22], 0.499906444 & 80 & 0.96474 & 0.96976 \\ \hline
\textbf{54} & 2.32228 & 2.97413 & 0.49990 & -0.22795569, -0.28454808 & 0.43776391, 0.06221192 & 80 & 0.96224 & 0.97109 \\ \hline
\textbf{55} & 2.58994 & 5.10622 & 0.32334 & -0.31939401, -0.0295772 & 0.00628181, 0.5 & 80 & 0.95194 & 0.97293 \\ \hline
\textbf{56} & 3.12728 & 2.66350 & 0.39019 & -0.632374964, -6.1994[-05] & -1.18584613[-20], 0.49994338 & 80 & 0.95171 & 0.96751 \\ \hline
\textbf{57} & 2.47000 & 5.43223 & 0.54154 & -0.50496711, -0.48283996 & 1.25185446[-32], 9.86319[-05] & 80 & 0.95510 & 0.96737 \\ \hline
\textbf{58} & 2.72399 & 3.71018 & 6.01040 & -0.16010373, -0.1469441 & 0.06064396, 0.5 & 80 & 0.95385 & 0.96758 \\ \hline
\textbf{59} & 2.90324 & 4.67714 & 1.73488 & -0.28429079, -0.20231022 & 0.00179217, 0.47709926 & 80 & 0.95081 & 0.97573 \\ \hline
\textbf{60} & 2.99424 & 4.13582 & 4.91289 & -0.02108512, -0.01182204 & 0.4291328, 0.49965538 & 80 & 0.95090 & 0.96142 \\ \hline
\textbf{61} & 2.41078 & 2.27029 & 2.80876 & -0.52971722, -2[-10] & 0, 0.4999877 & 80 & 0.95496 & 0.96923 \\ \hline
\textbf{62} & 2.04922 & 5.46014 & 5.29590 & -0.629668181, -1[-10] & 0.00678738, 0.49976816 & 80 & 0.95974 & 0.97135 \\ \hline
\textbf{63} & 2.21419 & 3.89046 & 5.83689 & -0.47849604, -0.23786693 & 0.01470454, 0.1980091 & 80 & 0.96645 & 0.97392 \\ \hline
\textbf{64} & 2.94148 & 1.41033 & 4.49399 & -0.91405141, -0.37960688 & 1.23144[-05], 0.015310517 & 80 & 0.95117 & 0.97334 \\ \hline
\textbf{65} & 2.15569 & 3.79110 & 0.80333 & -0.4722672, -0.0008256 & 2.7606[-05], 0.490354894 & 80 & 0.95660 & 0.96973 \\ \hline
\textbf{66} & 2.12855 & 6.09186 & 5.37105 & -0.68386331, -0.02070827 & 6.78906[-05], 0.445970143 & 80 & 0.95647 & 0.96928 \\ \hline
\textbf{67} & 2.46317 & 5.92169 & 0.92723 & -0.40536361, -0.06195646 & -9.71500658[-17], 0.5 & 80 & 0.95483 & 0.97081 \\ \hline
\textbf{68} & 2.57193 & 3.08012 & 2.47826 & -0.20387137, -0.15862918 & 3.9381[-05], 0.498799416 & 81 & 0.95168 & 0.97153 \\ \hline
\textbf{69} & 2.96948 & 0.51671 & 4.13995 & -0.3304364,   -0.25845847 & 0.16309126, 0.41210785 & 80 & 0.95115 & 0.96170 \\ \hline
\textbf{70} & 2.70184 & 0.12106 & 2.01037 & -0.38296615, -0.06624632 & 9.13122[-05], 0.499907461 & 80 & 0.95470 & 0.97134 \\ \hline
\textbf{71} & 2.66267 & 0.41965 & 3.55148 & -0.586690017, -0.000372t & 0.0002436, 0.499949076 & 80 & 0.95525 & 0.96859 \\ \hline
\textbf{72} & 2.98978 & 1.38898 & 1.96907 & -0.23519674, -0.21290984 & 0.12852131, 0.30134054 & 80 & 0.94986 & 0.96740 \\ \hline
\textbf{73} & 2.27941 & 5.09299 & 3.50890 & -0.23220527, -0.19757706 & 0.02101817, 0.44517371 & 80 & 0.95617 & 0.97342 \\ \hline
\textbf{74} & 2.91441 & 4.72570 & 0.86171 & -0.48678714, -0.00094883 & -3.74034123[-22], 0.5 & 80 & 0.95442 & 0.96971 \\ \hline
\textbf{75} & 2.75605 & 3.05189 & 4.90998 & -0.81900705, -0.33719561 & 0.00041039, 0.01351214 & 80 & 0.94987 & 0.97077 \\ \hline
\textbf{76} & 2.72950 & 3.56458 & 0.87768 & -0.40243751, -0.00236464 & 4.45289[-05], 0.499937424 & 80 & 0.95567 & 0.97289 \\ \hline
\textbf{77} & 2.76295 & 5.11002 & 0.06402 & -0.29332909, -0.1951465 & 0.0212047, 0.39766611 & 80 & 0.95471 & 0.97428 \\ \hline
\textbf{78} & 2.72094 & 1.39060 & 3.53269 & -0.82801909, -0.26676811 & 1.5504[-05], 0.201188085 & 80 & 0.95052 & 0.96384 \\ \hline
\textbf{79} & 2.99682 & 3.00681 & 2.64991 & -0.70675606, -0.01714809 & 3.9807[-05], 0.368567782 & 80 & 0.95371 & 0.96678 \\ \hline
\textbf{80} & 2.86026 & 4.39329 & 3.54436 & -0.372577823, -1[-10] & 6.98153[-08], 0.5 & 80 & 0.94878 & 0.97406 \\ \hline
\textbf{81} & 2.39940 & 0.21368 & 1.29938 & -0.45744, -0.1669 & -2.8281[-20], 0.38364 & 80 & 0.95497 & 0.97135 \\ \hline
\textbf{82} & 1.89477 & 2.81396 & 0.23890 & -0.510694, -0.092163 & -1.09101[-19], 0.5 & 80 & 0.96706 & 0.97563 \\ \hline
\textbf{83} & 2.18455 & 3.97272 & 1.39291 & -0.508156, -1[-10] & 0.00456, 0.49991 & 80 & 0.95857 & 0.96752 \\ \hline
\textbf{84} & 2.49180 & 2.58583 & 5.14075 & -0.99610258, -0.34906797 & 7.73614[-06], 6.6692[-05] & 80 & 0.95209 & 0.97350 \\ \hline
\textbf{85} & 2.59845 & 2.37573 & 3.12417 & -0.441353237, -6.2831[-05] & 0.01563258, 0.5 & 80 & 0.95124 & 0.97010 \\ \hline
\textbf{86} & 2.60883 & 6.17957 & 2.03449 & -0.21592729, -0.17875946 & 4.21849[-05], 0.312400669 & 80 & 0.95445 & 0.97418 \\ \hline
\textbf{87} & 2.85977 & 3.13459 & 4.87675 & -0.950751522, -1[-10] & 1.22478392[-18], 0.499001011 & 80 & 0.95092 & 0.96456 \\ \hline
\textbf{88} & 3.05267 & 2.09009 & 0.15869 & -0.553028459, -1[-10] & 1.41296829[-19], 0.409878614 & 80 & 0.94655 & 0.96793 \\ \hline
\textbf{89} & 2.65536 & 1.53336 & 2.01915 & -0.3610569,   -0.10599245 & 0.00771934, 0.49831975 & 80 & 0.95151 & 0.97561 \\ \hline
\textbf{90} & 2.42194 & 5.35202 & 1.63601 & -0.35453141, -0.32129783 & 0.12163206, 0.13424023 & 80 & 0.95596 & 0.96609 \\ \hline
\textbf{91} & 2.16767 & 1.66413 & 1.52262 & -0.98601261, -0.50422059 & -9.48289521[-21], 6.12274[-05] & 80 & 0.96180 & 0.97663 \\ \hline
\textbf{92} & 2.98966 & 3.05187 & 1.47806 & -0.529954412, -1.06161[-05] & 0.00924164, 0.49999361 & 80 & 0.95481 & 0.96844 \\ \hline
\textbf{93} & 2.80563 & 4.43216 & 0.93327 & -0.63947564, -0.01072315 & 4.78721[-05], 0.461841234 & 80 & 0.95126 & 0.96677 \\ \hline
\textbf{94} & 2.76116 & 3.67330 & 3.58557 & -0.28274626, -0.0566703 & -3.38813179[-20], 0.499436234 & 80 & 0.95401 & 0.97358 \\ \hline
\textbf{95} & 2.86817 & 4.82626 & 0.79992 & -0.3661714,   -0.32752026 & 0.11331283, 0.12163038 & 80 & 0.95044 & 0.96680 \\ \hline
\textbf{96} & 3.10749 & 6.05200 & 0.55580 & -0.30965446, -0.24969456 & 0.00307031, 0.15108169 & 80 & 0.95483 & 0.97068 \\ \hline
\textbf{97} & 2.52530 & 1.56729 & 5.60700 & -0.625100171, -2[-10] & 5.55111511[-21], 0.4999 & 80 & 0.95207 & 0.96550 \\ \hline
\textbf{98} & 2.61797 & 5.00611 & 4.61394 & -0.334751199, -1[-10] & 0.03123503, 0.5 & 80 & 0.95578 & 0.97218 \\ \hline
\textbf{99} & 2.53796 & 3.96390 & 1.80047 & -0.998360446, -2[-10] & 1.21919[-05], 0.48416587 & 80 & 0.95285 & 0.96310 \\ \hline
\textbf{100} & 2.31079 & 5.94527 & 1.46166 & -0.44647969, -0.27303986 & -3.88578059[-20], 0.150260118 & 80 & 0.95846 & 0.97214 \\ \hline
\caption{Data for learning accuracy.}
\label{table2}
\end{longtable*}

We can see in this case that the fidelity obtained with optimization increase respect to the fidelity obtained by the use of a uniform PDF, which means that the optimization of the learning accuracy works fine.


\begin{thebibliography}{999}
\bibitem{Russell1995Book} S. Russell and P. Norvig, \textit{Artificial Intelligence: A Modern Approach} (Prentice Hall, New York, 1995).

\bibitem{Michalski2013Book} R. S. Michalski, J. G. Carbonell, and T. M. Mitchell, \textit{Machine Learning: An Artificial Intelligence Approach} (Springer, Berlin, 2013).

\bibitem{Jordan2015Science} M. I. Jordan and T. M. Mitchell, \href{https://science.sciencemag.org/content/349/6245/255}{Science \textbf{349}, 255 (2015).}

\bibitem{Carleo2019} G. Carleo, I. Cirac, K. Cranmer, L. Daudet, M. Schuld, N. Tishby, L. Vogt-Maranto, and L. Zdeborov\'a, \href{https://journals.aps.org/rmp/abstract/10.1103/RevModPhys.91.045002#tr1}{Rev. Mod. Phys. {\bf 91}, 045002 (2019).}

\bibitem{Kaelbling1996JAIR} L. P. Kaelbling, M. L. Littman, and A. W. Moore, \href{https://www.jair.org/index.php/jair/article/view/10166}{J. Artif. Intell. \textbf{4}, 237 (1996).}

\bibitem{Sutton1998Book} R. S. Sutton and A. G. Barto, \textit{Reinforcement Learning: An Introduction} (MIT, Cambridge, 2018).
 
\bibitem{Silver2017Nature} D. Silver {\it et al.}, \href{https://www.nature.com/articles/nature24270?sf123103138=1}{Nature \textbf{550}, 354 (2017).}
 
\bibitem{Silver2018Science} D. Silver {\it et al.}, \href{https://science.sciencemag.org/content/362/6419/1140/}{Science \textbf{362}, 1140 (2018).}

\bibitem{Vinyals2019Nature} O. Vinyals {\it et al.}, \href{https://www.nature.com/articles/s41586-019-1724-z?}{Nature \textbf{575}, 350 (2019).}

\bibitem{Steane1998RPP} A. Steane, \href{https://iopscience.iop.org/article/10.1088/0034-4885/61/2/002/meta}{Rep. Prog. Phys. \textbf{61}, 117 (1998).}

\bibitem{Preskill2018Quantum} J. Preskill, \href{https://quantum-journal.org/papers/q-2018-08-06-79/}{Quantum \textbf{2}, 79 (2018)}

\bibitem{Gyongyosi2019CSR} L. Gyongyosi and S. Imre, \href{https://www.sciencedirect.com/science/article/abs/pii/S1574013718301709}{Comput. Sci. Rev. \textbf{31}, 51 (2019).}

\bibitem{Knill2010Nature} E. Knill, \href{https://www.nature.com/articles/463441a}{Nature \textbf{463}, 441 (2010).}

\bibitem{Haffner2008PhysRep} H. H\"affner, C. F. Ross, and R. Blatt, \href{https://www.sciencedirect.com/science/article/abs/pii/S0370157308003463}{Phys. Rep. \textbf{469}, 155 (2008).}

\bibitem{Benhelm2008NatPhys} J. Benhelm, G. Kirchmair, C. F. Roos, and R. Blatt, \href{https://www.nature.com/articles/nphys961}{Nat. Phys. \textbf{4}, 463 (2008).}

\bibitem{Bruzewicz2019APR} C. D. Bruzewicz, J. Chiaverini, R. McConnell, and   J. M. Sage, \href{https://aip.scitation.org/doi/abs/10.1063/1.5088164}{Appl. Phys. Rev. \textbf{6}, 021314 (2019).}

\bibitem{Mariantoni2011Science} M. Mariantoni {\it et al.}, \href{https://science.sciencemag.org/content/334/6052/61.abstract}{Science \textbf{334}, 61 (2011).}

\bibitem{Devoret2011Science} M. H. Devoret and R. J. Schoelkopf, \href{https://science.sciencemag.org/content/339/6124/1169.abstract}{Science \textbf{339}, 1169 (2011).}

\bibitem{Wendin2017RPP} G. Wendin, \href{https://iopscience.iop.org/article/10.1088/1361-6633/aa7e1a/meta}{Rep. Prog. Phys. \textbf{80}, 106001 (2017).}

\bibitem{Huang2020SCIS} H.-L. Huang, D. Wu, D. Fan, and X. Zhu, \href{https://link.springer.com/article/10.1007/s11432-020-2881-9}{Sci. China Inf. Sci. \textbf{63}, 180501 (2020).}

\bibitem{Slussarenko2019RPP} S. Slussarenko and G. J. Pryde, \href{https://aip.scitation.org/doi/full/10.1063/1.5115814}{Rep. Prog. Phys. \textbf{6}, 041303 (2019).}

\bibitem{Arute} F. Arute {\it et al.}, \href{https://www.nature.com/articles/s41586-019-1666-5}{Nature \textbf{574}, 505 (2019).}

\bibitem{WuPhysRevLett2021} Y. Wu {\it et al.}, \href{https://journals.aps.org/prl/abstract/10.1103/PhysRevLett.127.180501}{Phys. Rev. Lett. \textbf{127}, 180501 (2021)}

\bibitem{Zhong2020Science} H.-S. Zhong {\it et al.}, \href{https://science.sciencemag.org/content/370/6523/1460.abstract}{Science \textbf{370}, 1460 (2020).}

\bibitem{Schuld2015ContempPhys} M. Schuld, I. Sinayskiy, and F. Petruccione, \href{https://www.tandfonline.com/doi/abs/10.1080/00107514.2014.964942}{Contemp. Phys. \textbf{56}, 172 (2015).}

\bibitem{Biamonte2017Nature} J. Biamonte, P. Wittek, N. Pancotti, P. Rebentrost, N. Wiebe, and S. Lloyd, \href{https://www.nature.com/articles/nature23474}{Nature \textbf{549}, 195 (2017).}

\bibitem{Dunjko2018RPP} V. Dunjko and H. J. Briegel, \href{https://iopscience.iop.org/article/10.1088/1361-6633/aab406/meta}{Rep. Prog. Phys. \textbf{81}, 074001 (2018).}

\bibitem{Lamata2020MLST} L. Lamata, \href{https://iopscience.iop.org/article/10.1088/2632-2153/ab9803/meta}{Mach. Learn.: Sci. Technol. \textbf{1}, 033002 (2020).}

\bibitem{Harrow2009PRL} A. W. Harrow, A. Hassidim, and S. Lloyd, \href{https://journals.aps.org/prl/abstract/10.1103/PhysRevLett.103.150502}{Phys. Rev. Lett. \textbf{103}, 150502 (2009).}

\bibitem{Wang2017PRA} G. Wang, \href{https://journals.aps.org/pra/abstract/10.1103/PhysRevA.96.012335}{Phys. Rev. A \textbf{96}, 012335 (2017).}

\bibitem{Arrazola2019PRA} J. M. Arrazola, T. Kalajdzievski, C. Weedbrook, and S. Lloyd, \href{https://journals.aps.org/pra/abstract/10.1103/PhysRevA.100.032306}{Phys. Rev. A \textbf{100}, 032306 (2019).}

\bibitem{Xin2020PRA} T. Xin, S. Wei, J. Cui, J. Xiao, I. Arrazola, L. Lamata, X. Kong, D. Lu, E. Solano, and G. Long, \href{https://journals.aps.org/pra/abstract/10.1103/PhysRevA.101.032307}{Phys. Rev. A \textbf{101}, 032307 (2020).}

\bibitem{Lloyd2020arXiv} S. Lloyd, G. De Palma, C. Gokler, B. Kiani, Z.-W. Liu, M. Marvian, F. Tennie, and T. Palmer, \href{https://arxiv.org/abs/2011.06571}{arXiv:2011.06571 (2020).}

\bibitem{Bukov2018PRX} M. Bukov, A. G. R. Day, D. Sels, P. Weinberg, A. Polkovnikov, and P. Mehta, \href{https://journals.aps.org/prx/abstract/10.1103/PhysRevX.8.031086}{Phys. Rev. X \textbf{8}, 031086 (2018).}

\bibitem{Niu2019NPJ} M. Y. Niu, S. Boixo, V. N. Smelyanskiy, and H. Neven, \href{https://www.nature.com/articles/s41534-019-0141-3}{npj Quantum Inf. \textbf{5}, 33 (2019).}

\bibitem{An2019EPL} Z. An and D. L. Zhou, \href{https://iopscience.iop.org/article/10.1209/0295-5075/126/60002/meta}{EPL \textbf{126}, 60002 (2019).}

\bibitem{Wang2020PRL} Z. T. Wang, Y. Ashida, and M. Ueda, \href{https://journals.aps.org/prl/abstract/10.1103/PhysRevLett.125.100401}{Phys. Rev. Lett. \textbf{125}, 100401 (2020).}

\bibitem{AlbarranArriagada2018PRA} F. Albarr\'an-Arriagada, J. C. Retamal, E. Solano, and L. Lamata, \href{https://journals.aps.org/pra/abstract/10.1103/PhysRevA.98.042315}{Phys. Rev. A \textbf{98}, 042315 (2018).}

\bibitem{Yu2019AQT} S. Yu, F. Albarr\'an-Arriagada, J. C. Retamal, Y.-T. Wang, W. Liu, Z.-J. Ke, Y. Meng, Z.-P. Li, J.-S. Tang, E. Solano, L. Lamata, C.-F. Li, and G.-C. Guo, \href{https://onlinelibrary.wiley.com/doi/abs/10.1002/qute.201800074}{Adv. Quantum Technol. \textbf{2}, 1800074 (2019).}

\bibitem{Mackeprang2020QMI} J. Mackeprang, D. B. Rao Dasari, and J. Wrachtrup, \href{https://doi.org/10.1007/s42484-020-00016-8}{Quantum Mach. Intell. \textbf{2}, 5 (2020).} 

\bibitem{Bukov2018PRB} M. Bukov, \href{https://journals.aps.org/prb/abstract/10.1103/PhysRevB.98.224305}{Phys. Rev. B \textbf{98}, 224305 (2018).}

\bibitem{Zhang2020PRL} Y.-H. Zhang, P.-L. Zheng, Y. Zhang, and D.-L. Deng, \href{https://journals.aps.org/prl/abstract/10.1103/PhysRevLett.125.170501}{Phys. Rev. Lett \textbf{125}, 170501 (2020).}

\bibitem{Dong2008IEEE} D. Dong, C. Chen, H. Li, and T.-J. Tarn, \href{https://ieeexplore.ieee.org/document/4579244}{IEEE Trans. Sist. Man Cybern. B \textbf{38}, 1207 (2008).}

\bibitem{Paparo2014PRX} G. D. Paparo, V. Dunjko, A. Makmal, M. A. Martin-Delgado, and H. J. Briegel, \href{https://journals.aps.org/prx/abstract/10.1103/PhysRevX.4.031002}{Phys. Rev. X \textbf{4}, 031002 (2014).}

\bibitem{AlvarezRodriguez2014SciRep} U. Alvarez-Rodriguez, M. Sanz, L. Lamata, and E. Solano, \href{https://www.nature.com/articles/srep04910}{Sci. Rep. \textbf{4}, 4910 (2014).}

\bibitem{AlvarezRodriguez2016SciRep} U. Alvarez-Rodriguez, M. Sanz, L. Lamata, and E. Solano, \href{https://www.nature.com/articles/srep20956}{Sci. Rep. \textbf{6}, 20956 (2016).}

\bibitem{AlvarezRodriguez2018SciRep} U. Alvarez-Rodriguez, M. Sanz, L. Lamata, and E. Solano, \href{https://www.nature.com/articles/s41598-018-33125-3}{Sci. Rep. \textbf{8}, 14793 (2018).}

\bibitem{Patil2018SciRep} A. Patil, D. Saha, and S. Ganguly, \href{https://www.nature.com/articles/s41598-018-33125-3}{Sci. Rep. \textbf{8}, 14793 (2018).}

\bibitem{AlbarranArriagada2020MLST} F. Albarr\'an-Arriagada, J. C. Retamal, E. Solano, and L. Lamata, \href{https://iopscience.iop.org/article/10.1088/2632-2153/ab43b4/meta}{Mach. Learn.: Sci. Technol. \textbf{1}, 015002 (2020).}

\bibitem{Pan2020SciRep} C.-Y. Pan, M. Hao, N. Barraza, E. Solano, and F. Albarr\'an-Arriagada, \href{https://www.nature.com/articles/s41598-021-90534-7}{Sci. Rep. \textbf{11}, 12241 (2021).}

\bibitem{Peruzzo2014NatCom} A. Peruzzo, J. McClean, P. Shadbolt, M.-H. Yung, X.-Q. Zhou, P. J. Love, A. Aspuru-Guzik, and J. L. O'Brien , \href{https://www.nature.com/articles/ncomms5213}{Nat. Comm. \textbf{5}, 4213 (2014).}

\bibitem{McClean2016NJP} J. R. McClean, J. Romero, R. Babbush, and A. Aspuru-Guzik, \href{https://iopscience.iop.org/article/10.1088/1367-2630/18/2/023023}{New J. Phys. \textbf{18}, 023023 (2016).}

\bibitem{Ferguson2021PRL} R. R. Ferguson, L. Dellantonio, A. Al Balushi, K. Jansen, W. D\"ur, and C. A. Muschik, \href{https://journals.aps.org/prl/abstract/10.1103/PhysRevLett.126.220501}{Phys. Rev. Lett. \textbf{126}, 220501 (2021).}

\bibitem{HegadePhysRevApp2021}N. N. Hegade, K. Paul, Y. Ding, M. Sanz, F. Albarr\'an-Arriagada, E. Solano, and X. Chen, \href{https://journals.aps.org/prapplied/abstract/10.1103/PhysRevApplied.15.024038}{Phys. Rev. Appl. \textbf{15}, 024038 (2021).}

\bibitem{Steffen1990AstroA} M. Steffen, Astron. Astrophys. \textbf{239}, 443 (1990).

\end{thebibliography}
\end{document}